\documentclass[aps,pra,twocolumn,longbibliography,superscriptaddress,floatfix]{revtex4-1}

\usepackage{graphicx}
\usepackage{dcolumn}
\usepackage{bm}
\usepackage{amsmath}
\usepackage{mathtools}
\usepackage{graphicx}
\usepackage{ulem}
\usepackage{epstopdf}
\usepackage{subfigure}
\usepackage{color}
\usepackage{amsthm}
\usepackage{newlfont}
\usepackage{graphicx}
\usepackage{epstopdf}
\usepackage{appendix}
\usepackage[breaklinks=true]{hyperref}
\usepackage{breakcites}
\usepackage{textcomp}
\usepackage{appendix}
\usepackage{multirow}	
\usepackage{color}
\usepackage{amssymb}
\usepackage{epsfig}
\usepackage{bm}
\usepackage[american]{babel}
\usepackage{braket}
\usepackage{soul}
\usepackage{float}
\hypersetup{
     colorlinks   = true,
     linkcolor    = red,
     citecolor    = cyan,
     urlcolor     = magenta
     }

\begin{document}


\title{Wave-packet spreading in the disordered and nonlinear Su-Schrieffer-Heeger chain}

\author{Bertin Many Manda}
\affiliation{Laboratoire d’Acoustique de l’Universit\'e du Mans (LAUM), UMR 6613, Institut d'Acoustique - Graduate School (IA-GS), CNRS, Le Mans Universit\'e, France}

\author{Vassos Achilleos}
\affiliation{Laboratoire d’Acoustique de l’Universit\'e du Mans (LAUM), UMR 6613, Institut d'Acoustique - Graduate School (IA-GS), CNRS, Le Mans Universit\'e, France}

\author{Olivier Richoux}
\affiliation{Laboratoire d’Acoustique de l’Universit\'e du Mans (LAUM), UMR 6613, Institut d'Acoustique - Graduate School (IA-GS), CNRS, Le Mans Universit\'e, France}

\author{Charalampos Skokos}
\affiliation{Nonlinear Dynamics and Chaos group, Department of Mathematics and Applied Mathematics, University of Cape Town, Rondebosch, 7701 Cape Town, South Africa}

\author{Georgios Theocharis}
\affiliation{Laboratoire d’Acoustique de l’Universit\'e du Mans (LAUM), UMR 6613, Institut d'Acoustique - Graduate School (IA-GS), CNRS, Le Mans Universit\'e, France}

\date{\today}

\begin{abstract}
    We numerically investigate the characteristics of the long-time dynamics of a single-site wave-packet excitation in a disordered and nonlinear Su-Schrieffer-Heeger model. 
    In the linear regime, as the parameters controlling the topology of the system are varied, we show that the transition between two different topological phases is preceded by an anomalous diffusion, in contrast to Anderson localization 
    within these topological phases.
    In the presence of on-site nonlinearity this feature is lost due to  mode-mode interactions.
    Direct numerical simulations reveal that the characteristics of the asymptotic nonlinear wave-packet spreading are the same across the whole studied parameter space.
     Our findings underline the importance of  mode-mode interactions in nonlinear topological systems, which must be studied in order to define reliable nonlinear topological markers.
\end{abstract}

\maketitle


\section{\label{sec:intro}Introduction}

The study of topological insulators has received growing interest over the last decade. 
One of the characteristics of such materials is the support of localized waves at their
edges or interfaces, that are robust to defects or weak disorder.
Such robust wave transport has already been encountered in a vast number of materials both in theoretical studies and in experiments performed, among others, in optics~\cite{LJS2014,OPAGHLRSSZC2019,OPAGHLRSSZC2019,SB2021}, electronics~\cite{HK2010,QZ2011,E2018a,SLCK2020}, mechanics and acoustics~\cite{H2016,SH2016,MXC2019,XSHMY2020,HCH2021,SBPM2022}.
When the strength of disorder is sufficiently increased, 
the expected behavior is that localisation sets in and the system is driven to the Anderson-localized phase. This results to the halt of wave transport and disappearance of topological features. However, the interplay between topology and disorder is more complex.
Interesting enough, the reverse transition is also possible. Strong disorder can bring the system into a topologically non-trivial phase, to the so-called topological Anderson insulator (TAI) phase, and can lead to the emergence of protected edge states and quantised transport. The most successful experimental realisations of TAI phases involve engineered systems, like cold atomic gases~\cite{MADMHG2018}, photonic~\cite{SPLTLSRS2018,LYRXLHSPZCZ2020} and acoustic~\cite{ZF2020} crystals, and photonic quantum walks~\cite{LLXWYX2022}.

In many experimental situations, the presence of nonlinearity can also strongly alter the topological nature of systems (see~\cite{SLCK2020} and references therein).
These nonlinearities can, for instance, be rooted to Kerr-like effects in systems such as nonlinear optical waveguide arrays~\cite{ESMBA1998}, atomic Bose-Einstein condensate (BEC) in optical potentials~\cite{L2001,TS2001}, and synthetic momentum-state lattices~\cite{ASHLMZHG2021}.
In this context, the initial studies of the interactions between topology and nonlinearity have revealed a whole new realm of topological phenomena like topological (``self-induced") edge solitons~\cite{RZPLPDNSS2013,ACM2014,CT2019,JD2022,J2023}, unique gap solitons~\cite{LPRS2013} and domain walls~\cite{HVA2017,CT2019}. 

Research on topological physics focuses on ways to identify topological phases in various physical situations~\cite{HK2010,QZ2011,OPAGHLRSSZC2019}.
Regarding linear  periodic (clean) systems, the application of the band theory, especially the {\it bulk-edge correspondence}, connects the number of edge modes of a finite size system to the topological invariant of the gapped energy spectrum of its bulk counterpart~\cite{AOP2016}.
On the other hand, the most common method in experiments focuses on detecting topological features at the boundaries of finite size samples~\cite{KBFRBKADW2012,MAG2016,SGGLOLSBA2017}.
In the presence of disorder, where translational symmetry is broken, the band theory does not apply and thus the various topological invariants must be computed in the real space.
In situations where disorder is present, the Bott index~\cite{LH2010}, spectral localizer~\cite{L2015} and the local topological marker (LTM)~\cite{MADMHG2018} have proven to be robust topological indicators, which paved the way to the identification of TAI phases in several disordered topological systems, see e.g.~\cite{RB2014,MADMHG2018,SKCATY2021,SW2022}.
Moreover, some efforts have also been put into elaborating similar indices for nonlinear lattices~\cite{JD2022}.

Recently, the mean chiral displacement of the quench dynamics of an initial single-site wave-packet in the bulk of a (non-interacting) topological system was also introduced as an observable to detect its topological invariant, without the need of energy band filling or external field~\cite{CDDMPDDCSMLM2017,MADMHG2018,M2019}.
Consequently, this method is extremely versatile, having already found application in experimental scenarios involving clean~\cite{CDDMPDDCSMLM2017,WXQWYX2018,WLMGLTZJJ2019,XDXGCYY2020}, disordered~\cite{MADMHG2018} and driven~\cite{M2019,XDXGCYY2020} systems and numerical simulations, for example in disordered mechanical systems~\cite{SKCATY2021}.
Nevertheless, these experiments and numerical simulations often limit themselves to short-times, even when the standard statistical moments of the wave-packet are computed~\cite{CDDMPDDCSMLM2017}.
At the origin of this time limitation is the design of structures for experiments which are usually restricted to few unit cells while `long-time dynamics' means wider spreading of the wave-packet which involves large lattice sizes.
On the other hand, the numerical integration of a large lattice is, in general, a computationally demanding task. 
Consequently, the long-time dynamics of topological systems have not been properly tackled.
\begin{figure}
    \centering
    \includegraphics[width=\columnwidth]{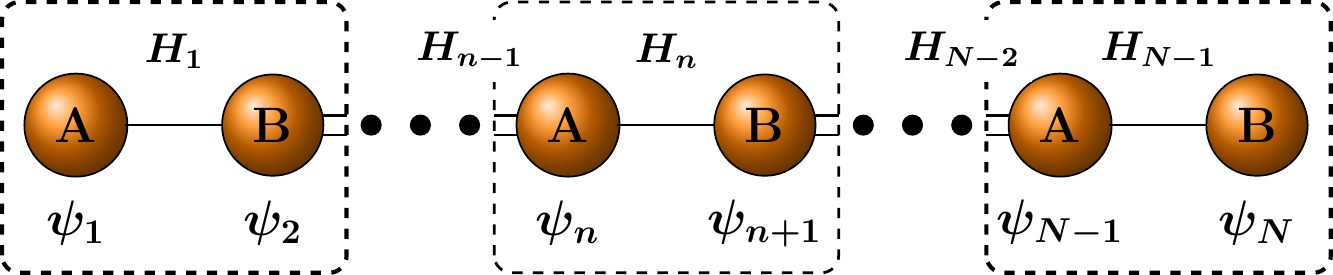}
    \caption{Schematic representation of the disordered SSH chain composed of $N$ sites.
    The chain is made up of two sublattices $A$ and $B$.
    Unit cells are highlighted with dashed boxes.
    The $\psi_n$ is the wave function at the $n$th site and $H_n$ is the hopping strength between the $n$th and $(n+1)$th sites. See text for details. 
    }
    \label{fig:ssh_chain}
\end{figure}

In this vein, a pertinent question is to know whether, and how, the mere observation of the long-time dynamics of an initially localized wave-packet in a disordered topological system is connected to the lattice's topological nature. 
The aim of this work is to tackle this problem, exploring the long-time dynamics of a single-site excitation wave-packet in systems supporting non-trivial topological phases in their linearized limits.
In particular, we focus on the disordered Su-Schrieffer-Heeger (SSH) lattice model~\cite{SSH1979}, the most famous one-dimensional (1D) lattice presenting topological features in its clean limit~\cite{AOP2016,BS2020}.
By varying the parameters controlling the topological phase of the chain, we track the characteristics of spreading of the wave-packet by computing the statistical moments of the amplitude distribution.
Our main finding is that, the transition between two regions with different winding numbers is characterized by an anomalous diffusion of the wave-packet whose moments grow pass a critical order, in contrast to its halt within the different topological phases for which all the moments saturate.

An additional aspect of our work is the extension of such studies to arbitrary strengths of on-site nonlinearity which, to the best of our knowledge, has not been addressed so far even in the monomer disordered SSH chain i.e. the 1D tight binding (TB) model with off-diagonal disorder~\cite{KL2011}.
Surprisingly, we find that the wave-packet spreading itself cannot indicate a topological transition, due to the presence of  mode-mode interactions.
More specifically, we show that at small and moderate nonlinearities, the wave-packet moments grow following the same power laws across the whole parameter space of the system.
On the other hand, when the nonlinearity is strong, the wave-packet spreading is partially or entirely  suppressed.

The paper is organized in the following way.
In Sec.~\ref{sec:review}, we briefly present the disordered SSH model and give a review of its spectral and topological properties.
In Sec.~\ref{sec:loc_length_finite_size}, we focus on the localization length of the modes with energies close to the band center along the topological transition curve.
In Sec.~\ref{sec:wave-packet_dynamics}, the spreading of single-site excitations is studied in the linear and nonlinear regimes.
Finally, we conclude our work in Sec.~\ref{sec:conclusion}.

\section{\label{sec:review}Overview of the properties of the disordered SSH chain}

\subsection{\label{subsec:spectral_properties}Spectral properties}

We start by reviewing some of the basic spectral properties of the 1D disordered SSH model (see also Fig.~\ref{fig:ssh_chain}) whose equation of motion for the complex wave function $\psi_n$ at the $n$th site of the chain reads,
\begin{equation}
    i\frac{d\psi_n}{dt} = H_{n} \psi_{n+1} + H_{n-1} \psi_{n-1},
    \label{eq:eq_motion_linear_general}
\end{equation}
with 
\begin{equation}
        H_{n-1} = 1 + W_2\epsilon_{n-1}, \mbox{ and }
        H_{n} = m + W_1\epsilon_n,
    \label{eq:hopping_general}
\end{equation}
where $m$ is the ratio between intra- and inter-cell hoppings in the clean (periodic) limit  (see Fig.~\ref{fig:ssh_chain}) and
$\epsilon_n$ are random numbers uniformly drawn from the interval $\left[-1/2, 1/2\right]$.
The $W_1$ and $W_2$ control the strengths of disorder on adjacent sites. Here, as in Refs.~\cite{MHSP2014,MADMHG2018}, we assume $W_1=2W_2=W$, 
allowing the reduction of the dimension of the system's parameter space, while allowing the appearance of the important physical phenomena we want to study.
It is worth noting that in this section of the parameter space, the hopping coefficients $H_n$ are positive in the region defined by the equation $W<\min \left\{4, 2m\right\}$.

For a chain of size $N$ ($N$ is even) the related set  equations of motion [Eq.~\eqref{eq:eq_motion_linear_general}] conserves both the total energy $\mathcal{H}_2$ and norm $\mathcal{A}$ of the system respectively
\begin{equation}
    \mathcal{H}_2 = \sum_{n=1}^{N} H_n( \psi_{n+1}^{\star} \psi_n +  \psi_{n+1}\psi_n^{\star}) ,\quad \mathcal{A} = \sum_{n=1}^{N} \lvert \psi_n\rvert^2.
    \label{eq:energy_norm_linear}
\end{equation}
Furthermore, unless otherwise stated, we set open boundary conditions at the two ends of the chain, i.e. $\psi_0 = \psi_{N+1} = 0$.

Let us first consider the related spectral problem. 
Substituting $\psi_{n} = B_n e^{-iEt}$ into the system's equations of motion [Eq.~\eqref{eq:eq_motion_linear_general}] leads to the eigenvalue problem 
\begin{equation}
    E \bm{B} = \mathbb{H}\bm{B},
    \label{eq:eigenvalue}
\end{equation}
where $E$ and $\bm{B} = (B_1, B_2, \ldots, B_N)^T$ [$\left(\cdot\right)^T$ denotes the matrix transpose] are respectively the system's energy and normal mode (NM) vector, with $\mathbb{H}$ being the Hamiltonian matrix of the system [the expression of this matrix is given in Eq.~\eqref{eq:Hamiltonian_matrix} of App.~\ref{app:sec:definitions}]. 
The Hamiltonian matrix $\mathbb{H}$ is tridiagonal with disorder appearing in its off-diagonal elements.
In addition $\mathbb{H}$ anti-commutes with the chiral operator $\Gamma$ [Eq.~\eqref{eq:chiral_matrices_01} of App.~\ref{app:sec:definitions}], i.e. 
\begin{equation}
    \Gamma\mathbb{H} + \mathbb{H}\Gamma = 0, \quad \forall \left(\epsilon_1, \epsilon_2, \ldots, \epsilon_N\right),
    \label{eq:chiral_symmetry}
\end{equation}
which means that it always possesses chiral symmetry.
This chiral symmetry is connected with the bipartite nature of the lattice~\cite{ITA1994}. This is the case when the lattice consists of two sublattices A and B, with nonzero hopping only between the A and B sites  (see~Fig.~\ref{fig:ssh_chain}).

A consequence of the chiral/bipartite symmetry of the disordered Hamiltonian matrix, is that its energy spectrum is symmetric about $E=0$.
In addition, as we show in App.~\ref{app:sec:zero_energy}, when the disordered SSH chain is in the topologically non-trivial region, it exhibits a pair of ``exact" zero states (in the thermodynamic limit). These states have wave functions (NMs) that vanish on one sublattice, see e.g.~\cite{IKM2012,MHSP2014,AOP2016}, and are most likely located around the two edges of the chain.
On the other hand, when the disordered SSH chain is in the topologically trivial region, it exhibits a pair of states with energy close to zero, for strong disorder. This pair of states now can be located anywhere along the chain.

\subsection{Topological properties}
\begin{figure}
    \centering
    \includegraphics[width=\columnwidth]{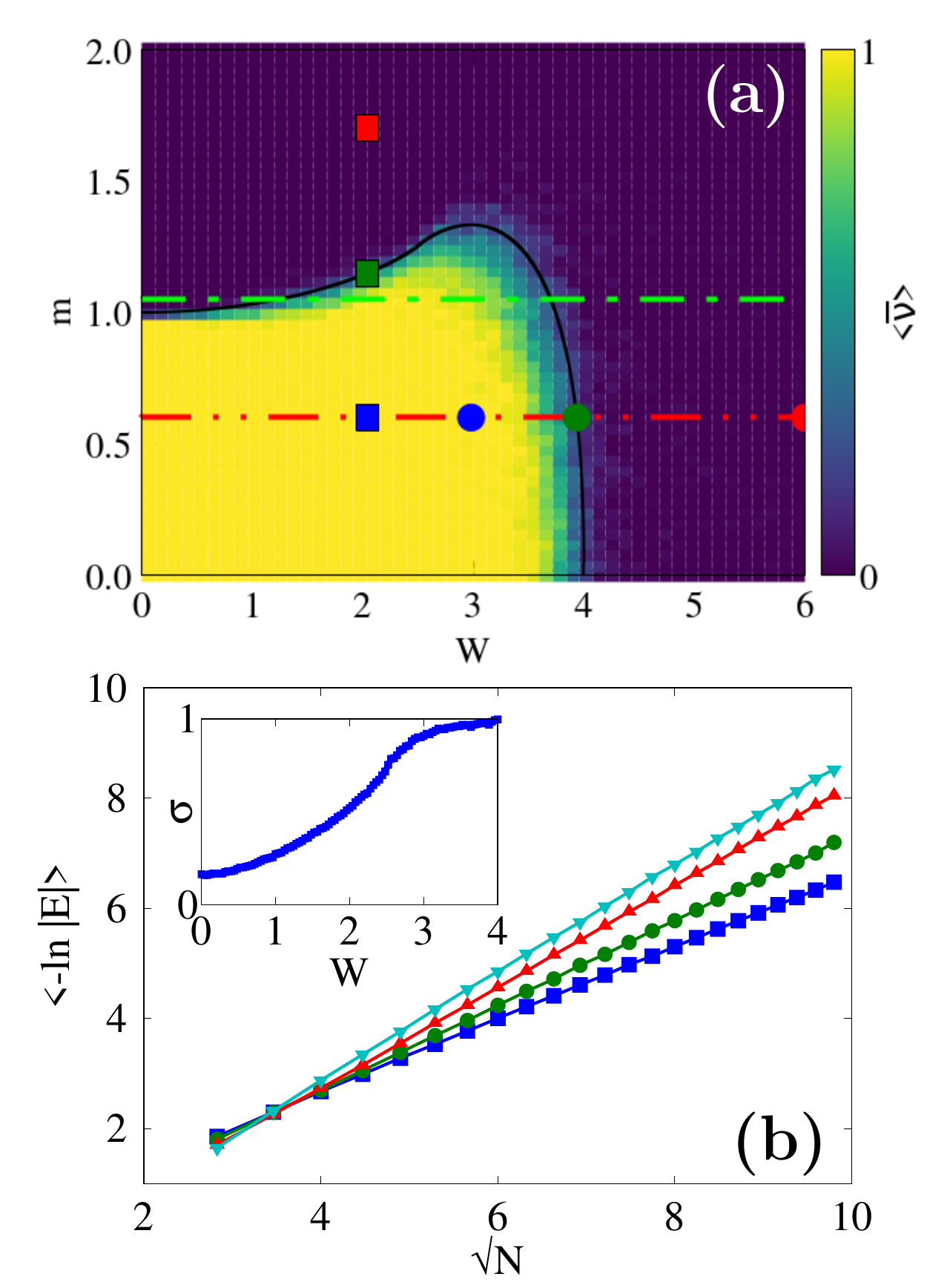}
    \caption{
    (a) Topological phase diagram of the disordered SSH chain [Eq.~\eqref{eq:eq_motion_linear_general}] generated by computing the winding number, $\langle \overline{\nu}\rangle$, of the chain in the $(W, m)$ parameter space.
    Random positive values of all the hoppings correspond to the region defined by $W < \min_{m} \left\{4, 2m\right\}$.
    The dashed-dotted-dotted red and dashed-dotted green lines represent sections of the phase diagram respectively at $m=0.6$ and $m=1.05$.
    Moreover the blue, green and red squares represent $(W=2.04, m=0.6)$, $(W=2.04, m=1.15)$, and $(W=2.04, m=1.7)$ respectively.
    On the other hand, the blue, green and red circles stand for $(W=3, m=0.6)$, $(W=3.94, m=0.6)$ and $(W=6, m=0.6)$.
    (b) $\langle-\ln \lvert E \rvert\rangle$ as function of $\sqrt{N}$ for four sets of parameters along the topological transition curve [black curve in (a)] with $(W=2.36, m=1.21)$ (blue line-connected squares), $(W=2.56, m=1.27)$ (green line-connected circles), $(W=3.00, m=1.33)$ (red line-connected triangles) and $(W=3.88, m=0.77)$ (cyan line-connected inverse-triangles). 
    Inset: Dependence of the slope $\sigma$ of the fitting functional form $\langle -\ln \lvert E \rvert\rangle \approx \sigma \sqrt{N}$ with respect to $W$.}
    \label{fig:topo_phase_diagram}
\end{figure}

A way to define a winding number for the disordered SSH chain, and in general for any chiral symmetric lattice, is through the LTM. 
In fact, the standard definiton of the winding number is expressed in momentum space of periodic lattices. 
This winding number can be rewritten in terms of the position operator in real space~\cite{BR2011,BR2013}. 
The latter could then be worked out per unit cell (or volume), such that a topological marker can be evaluated locally. 
For the finite disordered SSH chain with open boundary conditions, it is evaluated as follows.

Considering a fixed configuration of random hopping strengths, we can diagonalize the above problem [Eq.~\eqref{eq:eigenvalue}] and obtain $N$ energies $E_\mu$ associated with $N$ normalized eigenvectors $\bm{B}_\mu = (B_{1,\mu}, B_{2,\mu}, \ldots, B_{N,\mu})^{T}$ $(\sum_{n=1}^{N} \lvert B_{n,\mu}\rvert^2= 1)$ which we sort in ascending order of energies.
We then form the matrices $\mathbb{B}_{-} = (\bm{B}_1, \bm{B}_2, \ldots, \bm{B}_{\frac{N}{2}-1}, \bm{B}_{\frac{N}{2}})$ and $\mathbb{B}_{+}=(\bm{B}_{\frac{N}{2}+1}, \bm{B}_{\frac{N}{2}+2}, \ldots, \bm{B}_{N-1}, \bm{B}_{N})$ and define the projectors on the positive, and negative energy bands respectively
$\mathbb{P}_{+}=\mathbb{B}_{+}\mathbb{B}_{+}^{T}$ and $\mathbb{P}_{-}=\mathbb{B}_{-}\mathbb{B}_{-}^{T}$.
Consequently, we write the matrix
$\mathbb{Q} = \mathbb{P}_{+} - \mathbb{P}_{-}$ which factors as $\mathbb{Q}=\mathbb{Q}_{AB}+\mathbb{Q}_{BA} = \Gamma_A \mathbb{Q} \Gamma_B + \Gamma_B \mathbb{Q} \Gamma_A$, with $\Gamma_A$ and $\Gamma_B$ being the projector operators to the $A$ and $B$ sublattices respectively [see Eq.~\eqref{eq:chiral_matrices_01} of App.~\ref{app:sec:definitions}].
After implementing the above steps, we calculate the LTM, $\nu(l)$~\cite{MHSP2014,MADMHG2018}, at the unit cell of index $l=1, 2, \ldots, N/2$ 
\begin{equation}
    \begin{split}
        \nu (l) &=\sum_{k=\{A,B\}}
        \frac{\left\{\left(\mathbb{Q}_{BA} \left[\mathbb{X}, \mathbb{Q}_{AB}\right] \right)_{lk,lk} + \left(\mathbb{Q}_{AB} \left[\mathbb{Q}_{BA}, \mathbb{X}\right] \right)_{lk,lk}
        \right\}}{2},
    \end{split}
    \label{eq:local_top_marker}
\end{equation}
with $\mathbb{X}$ being the position matrix [Eq.~\eqref{eq:position_matrix} in App.~\ref{app:sec:definitions}], 
and $lA$ ($lB$) referring to the elements of the matrix belonging to the $A$ ($B$) sublattice of the $l$th unit cell (Fig.~\ref{fig:ssh_chain}).
In practice, we perform a space-and-disorder average of $\nu$ [Eq.~\eqref{eq:local_top_marker}] to obtain 
the winding number, $\langle \overline{\nu} \rangle$, of the disordered chain~\cite{MHSP2014}.
Note that in the following, we denote by $\overline{\mathcal{Q}} = \mathcal{D}^{-1}\int_{\mathcal{D}} \mathcal{Q}d\mathcal{D}$ any space or time average of the observable $\mathcal{Q}$ over the domain $\mathcal{D}$, and $\langle \mathcal{Q}\rangle$ corresponds to the average of the same observable across several configurations of disorder.

Figure~\ref{fig:topo_phase_diagram}(a) depicts the topological phase diagram of the disordered SSH chain through the calculation of the 
lattice winding number, $\langle \overline{\nu} \rangle$, obtained by averaging the LTM [Eq.~\eqref{eq:local_top_marker}] across $25$ central cells and about $250$ configurations of disorder
throughout the $(W, m)$ parameter space. 
Note that for all the calculations showed in Fig.~\ref{fig:topo_phase_diagram}(a), a lattice size of $N=500$ sites was used to mimic the thermodynamic limit and diminish finite size effects.
The phase diagram maps two regions in the $(W,m)$ parameter space: a topologically non-trivial region [yellow colored area in Fig.~\ref{fig:topo_phase_diagram}(a)] for which the winding number is close to unity, i.e.~$\langle \overline{\nu}\rangle \approx 1$, encircled by a topologically trivial region [blue colored area in Fig.~\ref{fig:topo_phase_diagram}(a)] with practically zero winding number, i.e.~$\langle \overline{\nu}\rangle \approx 0$.

These two topological regions are well separated by a solid black curve. This curve indicates the set of critical points where the localization length, $\zeta$, (considering ordinary exponential localization, $\psi_n \approx e^{-\lvert n\rvert/\zeta}$) at zero energy diverges in the thermodynamic limit. As it was shown in Ref.~\cite{MHSP2014}, this takes the following analytical form:
\begin{equation}
    \zeta (E=0) \approx \left\lvert  
        \ln \left[\frac{
            \left\lvert 4 - W \right\rvert^{\frac{2}{W} - \frac{1}{2}}
            \left\lvert 2m + W \right\rvert^{\frac{m}{W} + \frac{1}{2}}
        }{
            \left\lvert 4 + W \right\rvert^{\frac{2}{W} + \frac{1}{2}} 
            \left\lvert 2m - W \right\rvert^{\frac{m}{W} - \frac{1}{2}}
        } \right] \right\rvert.
    \label{eq:loc_length_zero_energy}
\end{equation}

\section{\label{sec:loc_length_finite_size}Localization length along the transition curve}

For $m=1$, the disordered SSH is reduced to the monomer TB lattice with off-diagonal disordered hoppings. As it is known~\cite{FL1977,SE1981,IKM2012}
in these lattices (and in general disordered bipartite lattices ~\cite{ITA1994}) the zero-energy wave function follows an unusual localization property. In particular,  it follows the form 
\begin{equation}
   \ln\lvert \psi_n \rvert \approx \sigma \sqrt{n},
\end{equation}
where $\sigma$ is a real scaling coefficient.
This wave function scaling explains the divergence of the
ordinary exponential localization of Eq.~\eqref{eq:loc_length_zero_energy}.

It is also known~\cite{D1953,KL2011,IKM2012} that the localization length, $\zeta (E)$, for small energies and in the thermodynamic limit takes the following form  
\begin{equation}
    \zeta (E) \approx \frac{-\ln \lvert E \rvert }{\sigma^2}.
    \label{eq:loclength_dyson_01}
\end{equation}
Thus it is implied that for a finite lattice of size $N$ the following relation holds
\begin{equation}
    \langle-\ln \lvert E \rvert\rangle \approx \sigma \sqrt{N}.
    \label{eq:scaling_coeff}
\end{equation}
The above behavior is known for the case of $m=1$~\cite{KL2011}. It 
is thus natural to ask if the same scaling of the energy close to  $E=0$ is valid all along the topological transition curve [black line in Fig~\ref{fig:topo_phase_diagram}(a)] and the answer is affirmative. In fact, by calculating the energy of the two modes closest to zero~\cite{NOTE2022a} for different lattice sizes (averaged over $10^4$ configurations of disorder), we have confirmed that the relation of Eq.~\eqref{eq:scaling_coeff} holds along the topological transition curve. Four typical examples are shown in Fig.~\ref{fig:topo_phase_diagram}(b).
It is clear that $\langle-\ln \lvert E\rvert\rangle$ grows linearly with $\sqrt{N}$ for all cases. Furthermore by using a linear fit we obtain the value of the parameter $\sigma$ for $100$ sets of parameters $(W, m)$ and the outcomes of this process is shown in the inset of Fig.~\ref{fig:topo_phase_diagram}(b).
We clearly see that $\sigma$ is close to zero at small $W$, while it grows as $W$ increases, tending to saturate toward $\sigma\approx 1$ for $W\rightarrow 4$.

\section{\label{sec:wave-packet_dynamics}Wave-packet dynamics}

After reviewing, the spectral properties and presenting the topological phase diagram of the disordered SSH chain, we investigate in this section the dynamics of an initially localized wave-packet. In particular, we are interested in the spreading of the wave-packet  in the $(W, m)$ parameter space.
Thus, we follow the time evolution of an initial wave-packet located on a single-site at the center of the lattice
\begin{equation}
    \psi_{n} (t=0) = \left\{ 
    \begin{split}
        \sqrt{a}e^{i\phi}, &\quad \mbox{if $n=\frac{N}{2}+1$}\\
        0\qquad, &\quad \mbox{otherwise}
    \end{split}
     \right.
     \label{eq:initial_wave_packet}
\end{equation}
with $a = 1$ the norm at site $N/2+1$, and $\phi = \pi/2$ its phase. 
In the rest of this work, the equations of motion [Eq.~\eqref{eq:eq_motion_linear_general}] of the linear (nonlinear) disordered SSH chain are integrated using the symplectic $ABA864$ ($s11ABC6$) scheme~\cite{HWL2006,BCFLMM2013,SGBPE2014,GMS2016,DMMS2019} and we monitor the accuracy of our simulations by evaluating the relative energy and norm errors,  $E_r(t) = \lvert \left(\mathcal{H}(t) - \mathcal{H}(0)\right)/\mathcal{H}(0)\rvert$ and $A_r(t) = \lvert \left(\mathcal{A}(t) - \mathcal{A}(0)\right)/\mathcal{A}(0)\rvert$, respectively. 
The use of an integration time step $\tau\approx 0.04-0.2$ ensures that the $E_r(t)$ and $S_r(t)$ are always bounded from above by $10^{-4}$ in all our simulations.
Unless otherwise stated, our lattice size is fixed at $N=10^4$ sites, a number which is sufficient to secure the avoidance of finite size effects up to integration times $t\approx 10^4-10^7$. 

As basic observables of the wave spreading dynamics, we compute the spatial moments of the averaged amplitude distribution at time $t>0$~\cite{LSA2010,KL2011,ATS2016}
\begin{equation}
    M_q (t) =\left\langle \sum_{l=1}^{L} \lvert l - l_0\rvert^q \chi _l(t) \right\rangle,~\chi_l(t) =  \frac{1}{\mathcal{A}} \sum_{k=\{ A, B\}} \lvert \psi_{lk}(t) \rvert^2,
    \label{eq:moments_of_evolution}
\end{equation}
with $q$ being the moment order, $l_0$, the cell index of the initially excited site, $L=N/2$ the total number of cells, $\chi_l$ the normalized norm at the cell of index $l$ and  $\mathcal{A}$ [Eq.~\eqref{eq:energy_norm_linear}] the total norm of the system.
In the present work, we focus on the first four integer orders of the moment, i.e. $M_q$ with $q=1,2,3,4$, which are sufficient to infer the spatial dynamics of the wave-packet.

\subsection{\label{subsec:wave-packet_linear}Linear limit}

\begin{figure*}
    \centering
    \includegraphics[width=\textwidth]{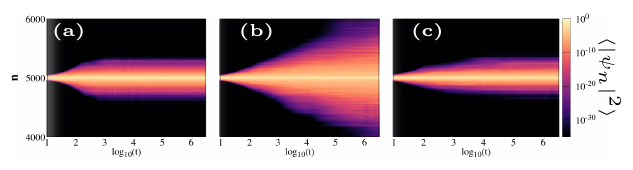}
    \caption{
            Time evolution of the averaged amplitude distribution, $\langle \lvert \psi_n \rvert^2 \rangle$, for three typical sets of parameters along the line $m=0.6$ with (a) $W=3.0$ inside the topologically trivial phase, (b) $ W=3.94$ at the topological transition and (c) $W=6.0$ inside the topologically non-trivial region [these points are respectively represented by the blue, green and red circles in Fig.~\ref{fig:topo_phase_diagram}(a)].
        In all panels, points are colored according to the magnitude of the wave-packet's amplitude by using the color scale at the right side of the figure.
      The amplitudes are averaged over $2000$ configurations of disorder.
    }
    \label{fig:norm_W_against_time_m_constant}
\end{figure*}

We start by qualitatively characterizing the spreading dynamics of the disordered SSH chain in the $(W, m)$ space.
In Fig.~\ref{fig:topo_phase_diagram}(a) we show three typical sets of parameters along the $m=0.6$ line [red dashed-dotted-dotted line in Fig.~\ref{fig:topo_phase_diagram}(a)] with $W=3$ [blue circle], $W=3.94$ [green circle]  and $W=6$ [red circle].
The first and the last cases respectively belong to the topologically non-trivial and trivial regions of the $(W,m)$ space, while the second one maps on the topological transition curve.
For these sets of parameters, we present in Fig.~\ref{fig:norm_W_against_time_m_constant} the time evolution of the amplitude distributions in real space averaged over $2000$ configurations of disorder.
All simulations in Fig.~\ref{fig:norm_W_against_time_m_constant} are carried up to $t\approx 10^6$.

Clearly, for the representative sets of parameters inside the topologically trivial and non-trivial regimes, we see that the wave-packets tend to spread away from the initial site of excitation at the early stage of the evolution followed by a tendency to saturate at large times [Figs.~\ref{fig:norm_W_against_time_m_constant}(a) and (c)].
This behavior hints toward Anderson localization (AL)~\cite{A1958,KM1993}.
On the other hand, for the case at the topological transition in Fig.~\ref{fig:norm_W_against_time_m_constant}(b), we observe
that, although the disorder is stronger than the one of Fig.~\ref{fig:norm_W_against_time_m_constant}(a), the  wave-packet surprisingly appears to spread throughout the lattice.

Let us now give an explanation to the wave-packet dynamical features observed above. 
Following Ref.~\cite{KL2011}, we can rewrite the moments $M_q$ [Eq.~\eqref{eq:moments_of_evolution}] in the limit of a large lattice as
\begin{equation}
    M_q \propto \int_{E} \zeta^q (E) \rho (E) dE,
    \label{eq:moments_analytical_01}
\end{equation}
with $\zeta(E)$ and $\rho(E)$ being the density of states and the localization length respectively, which depend on the energy $E$, but also on the parameters $(W, m)$ (see Secs.~\ref{sec:review} and~\ref{sec:loc_length_finite_size}).
It is worth emphasizing that the integration in Eq.~\eqref{eq:moments_analytical_01} is carried over the whole energy band and is valid everywhere in the $(W, m)$ parameter space of Fig.~\ref{fig:topo_phase_diagram}(a).
Then, in the topologically trivial and non-trivial phases our numerical computations of the localization length showed that this quantity is bounded from above across the whole energy spectrum due to AL, i.e. $\zeta(E) \leq \zeta_{max}$
(see also Ref.~\cite{MHSP2014}).
Thus Eq.~\eqref{eq:moments_analytical_01} gives 
\begin{equation}
    M_q (t) \propto \zeta_{max}^q,
    \label{eq:moments_analytical_Anderson}
\end{equation}
i.e. a saturation of the wave-packet moments within both phases, in agreement with the  results presented in Fig.~\ref{fig:norm_W_against_time_m_constant}(a) and  Fig.~\ref{fig:norm_W_against_time_m_constant}(c).

\begin{figure}
    \centering
    \includegraphics[width=\columnwidth]{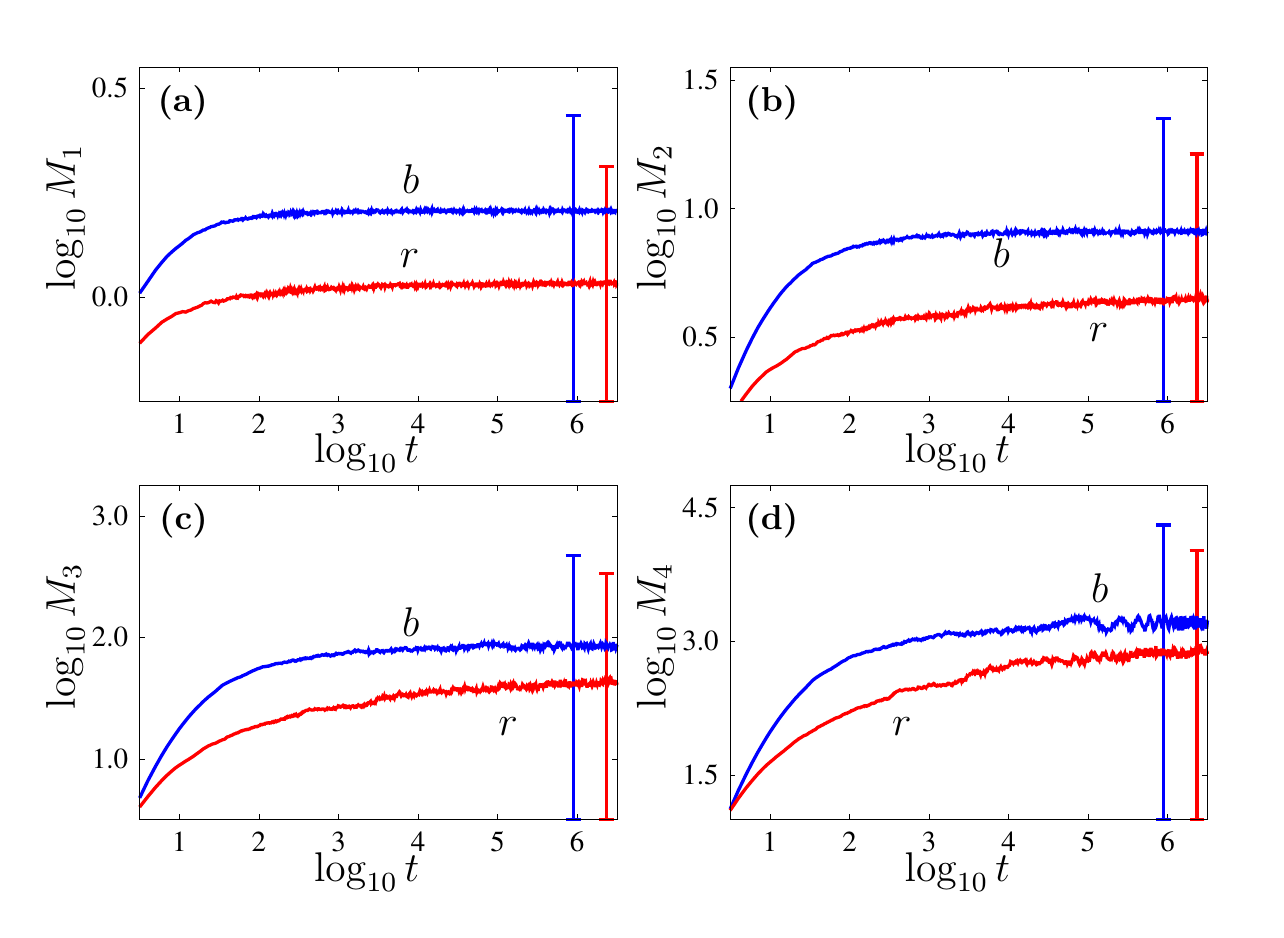}
    \caption{
        Time evolution of the moments  (a) $M_1$,  (b) $M_2$,  (c) $M_3$ and  (d) $M_4$  [Eq.~\eqref{eq:moments_of_evolution}], averaged over $9000$ configurations of disorder,
        for the disordered SSH chain of Eq.~\eqref{eq:eq_motion_linear_general}.
        The blue (b) and red (r) colored curves correspond to $(W=3, m=0.6)$ and $(W=6., m=0.6)$ respectively.
        The error bars of the same color as the curves, denote one standard deviation.
    }
    \label{fig:Mq_m_constant}
\end{figure}

To confirm Eq.~\eqref{eq:moments_analytical_Anderson} we performed extensive numerical simulations of the propagation of an initially localized wave-packet for the two points along the line $m=0.6$ with $W=3$ of Fig.~\ref{fig:norm_W_against_time_m_constant}(a) and $W=6$ of Fig.~\ref{fig:norm_W_against_time_m_constant}(c).
In Fig.~\ref{fig:Mq_m_constant}, we present the time evolution of the moments $M_q$, $q=1,2,3,4$ of the wave-packet averaged over $9000$ configurations of disorder for these two cases.
For both cases, the $M_1(t)$ [Fig.~\ref{fig:Mq_m_constant}(a)], $M_2(t)$ [Fig.~\ref{fig:Mq_m_constant}(b)], $M_3(t)$ [Fig.~\ref{fig:Mq_m_constant}(c)] and $M_4(t)$ [Fig.~\ref{fig:Mq_m_constant}(d)] tend to grow at the early stage of the evolution, corresponding to the time the wave-packet expands over the localization length of all excited modes [see also Fig.~\ref{fig:norm_W_against_time_m_constant}(a) and Fig.~\ref{fig:norm_W_against_time_m_constant}(c)].
At large times however, the moments asymptotically saturate to values that are different for each case.
Indeed, we see that for $W=3$ [blue curves in Fig.~\ref{fig:Mq_m_constant}] the moments levels off  to values larger than the ones seen in case of $W=6$ [red curves in Fig.~\ref{fig:Mq_m_constant}].
These observations are in agreement with the predictions of Eq.~\eqref{eq:moments_analytical_Anderson}, since the localization length is in general larger for smaller $W$ values. 
In addition, the large error bars seen in Fig.~\ref{fig:Mq_m_constant} are due to the presence of upper outliers of $M_q(t)$ originating from configurations of disorder for which the initial wave-packet excites mostly the modes close to the zero energy.
The localization lengths of these modes are much larger than the averaged localization length of the system, reminiscent of the Dyson singularity.
The existence of such cases substantially contributes to the moments' mean value $M_q$ which is obtained through arithmetic average and increases the computed error (standard deviation).
Furthermore, in App.~\ref{app:sec:wave_packet_spreading}, we consider other sets of parameters within the topologically trivial and non-trivial regions to emphasize the generality of these findings.

On the other hand, the peculiar spreading at the topological transition curve can be explained as follows. As inferred by the results of Fig.~\ref{fig:topo_phase_diagram}(b) for $E\rightarrow 0$, at the transition curve, the localization length, $\zeta(E)$,  of the disordered SSH model shows the same behavior as the 1D TB with off-diagonal disorder, namely
$\zeta(E) \approx -\ln \lvert E \rvert /\sigma^2$ [Eq.~\eqref{eq:loclength_dyson_01}]. In addition, according to Ref.~\cite{KB2021}, this implies that the density of states 
$\rho (E)$ also diverges when $E\rightarrow 0$, following $\rho(E) = \sigma^2/(-\lvert E \rvert\ln^3 \lvert E\rvert)$.
As a consequence, we can directly apply the results of
Ref.~\cite{KL2011}, and find an  approximate analytical expression for the integral of Eq.~\eqref{eq:moments_analytical_01} yielding the following asymptotic logarithmic law for the moments
\begin{equation}
    M_q(t) \propto \left\{
    \begin{split}
            \frac{1}{\sigma^{2(q-1)}} \left[\ln t \right]^{2(q-2)}, & \qquad \mbox{if $q > 2$,} \\ 
            \frac{1}{\sigma^2} \ln \left[\frac{1}{\sigma^2} \left(\ln t \right)^2 \right], & \qquad \mbox{if $q=2$} \\ 
            cst/\sigma^2 , &\qquad \mbox{if $q<2$.}
    \end{split}
    \right..
    \label{eq:moments_KL}
\end{equation}
The above equations dictate that the moments with $q<2$ saturate to a constant value while all the moments with $q\ge 2$ diverge with time following a  logarithmic rate.
In the above expressions, $\sigma$ is the scaling coefficient of Eq.~\eqref{eq:loclength_dyson_01} which was numerically calculated using Eq.~\eqref{eq:scaling_coeff} [inset of Fig.~\ref{fig:topo_phase_diagram}(b)].



\begin{figure*}
    \centering
    \includegraphics[scale=1]{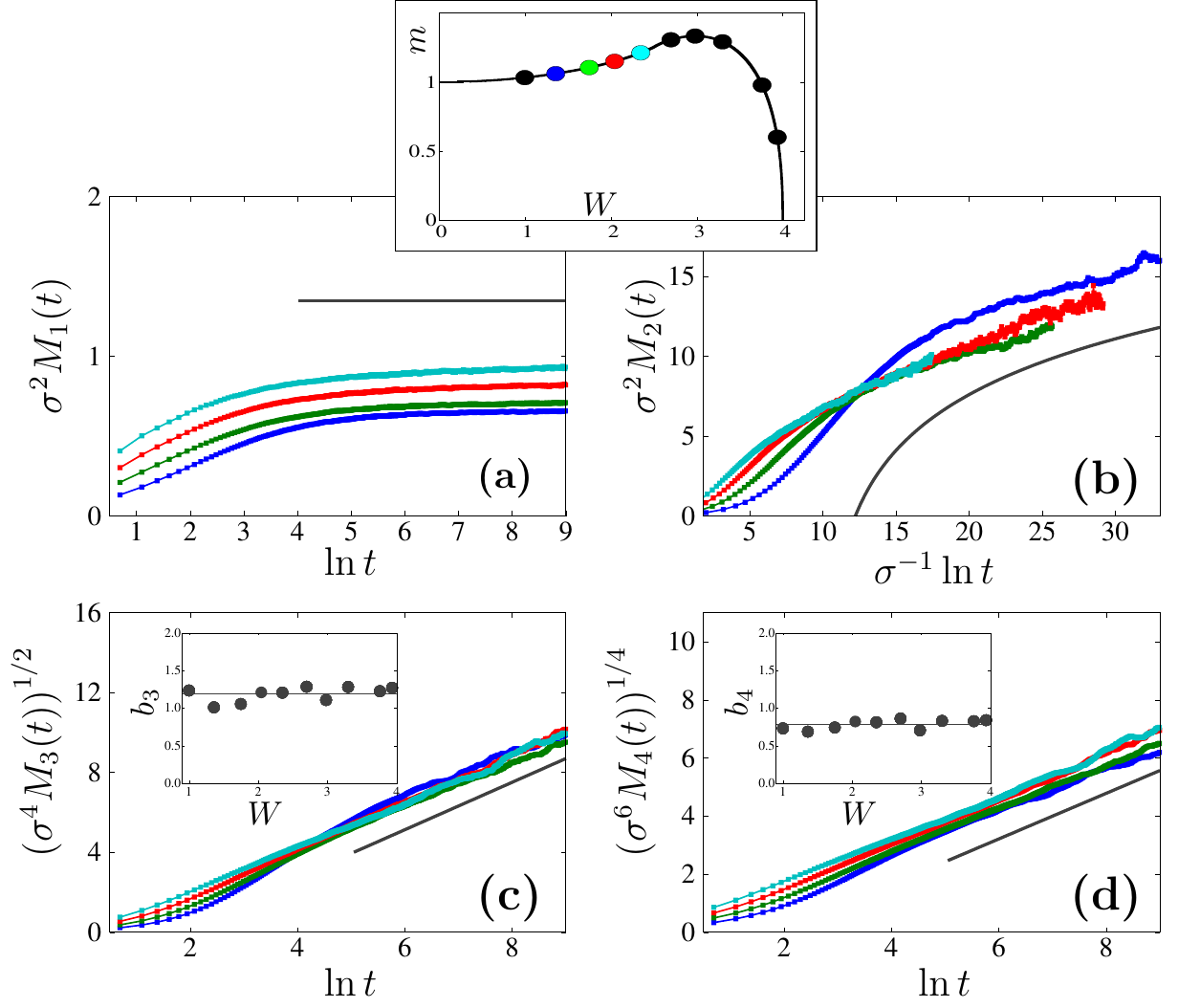}
    \caption{Dependence of 
    (a) $\sigma^2M_1(t)$ and 
    (b) $\sigma^2 M_2(t)$  
    (c) $(\sigma^4 M_3(t))^{1/2}$ and (d) $(\sigma^6M_4(t))^{1/4}$ on $\ln t$, $\ln t /\sigma$, $\ln t$ and $\ln t$ respectively
    for several sets of parameters along the topological transition line, depicted in the top panel. 
    The moments are averaged over $9000$ configurations of disorder.
    In addition, $\sigma$ in each case was numerically computed and reported in the inset of Fig.~\ref{fig:topo_phase_diagram}(b).
    The black straight lines in panels (a), (c) and (d) guide the eye for slopes (a) $0$, (c) $1.17$ [black line in the inset of (c)] and (d) $0.76$ [black line in the inset of (d)].
    Furthermore, the black curve in (b) corresponds to the logarithmic law $2.52\times \ln \left[ (\ln t/\sigma)^2\right]$.
    Insets: Dependence of the coefficients (c) $b_3$ [$(\sigma^4 M_3(t))^{1/2} \approx b_3 (\ln t)$] and (d) $b_4$ [$(\sigma^6M_4(t))^{1/4} \approx b_4 (\ln  t)$] on the disorder strength $W$.
    }
    \label{fig:KL_comparision_M3M4}
\end{figure*}

In order to support the analysis above, we perform numerical simulations of the propagation of an initially localized wave-packet [Eq.~\eqref{eq:initial_wave_packet}], by considering $10$ sets of parameters along the topological transition curve represented by the points in the phase diagram at the top panel of Fig.~\ref{fig:KL_comparision_M3M4}.
For these cases we compute the first four integer moments of spreading of the wave-packet, averaged over $9000$ configurations of disorder up to time $t\approx 10^4$.
In Fig.~\ref{fig:KL_comparision_M3M4}(a), we plot the results of the time evolution of the $\sigma^2 M_1(t)$ as function of $\ln t$ for some of these cases
(colored points in the phase diagram at the top panel of Fig.~\ref{fig:KL_comparision_M3M4}).
A saturation of the rescaled first moment to practically constant values at large time is clearly visible for all cases.
We find that for all the $10$ sets of parameters ranging within $W\in [1, 4]$ the constant values of $\sigma^2 M_1 \in [0.5, 1.5]$ at the final time of integration. 
Furthermore, these asymptotic constant values tend to grow with increasing disorder strength $W$.

In addition, in Fig.~\ref{fig:KL_comparision_M3M4}(b), we plot $\sigma^2M_2(t)$ as a function of $\ln t/\sigma$ for the same set of initial conditions as above.
The rescaled second moments grow in time for all cases, tending toward the same asymptotic law of evolution at large times.
In Fig.~\ref{fig:KL_comparision_M3M4}(b), we superimpose onto the time evolution of $\sigma^2M_2(t)$ the curve $b_2\ln \left[(\ln t/\sigma)^2\right]$ of Eq.~\eqref{eq:moments_KL} with $b_2$ being an arbitrary prefactor we fix to $b_2 \approx 2.52$ and find a good agreement with the numerical results.

Furthermore, according to the analytical expression for the moments' evolution [Eq.~\eqref{eq:moments_KL}], we expect for example that $(\sigma^4 M_3(t))^{1/2}$ and $(\sigma^6M_4(t))^{1/4}$ depend linearly on $\ln t$.
This assumption can be easily checked by numerically finding two constant real numbers, $b_3$ and $b_4$, such that
\begin{equation}
    (\sigma^4 M_3(t))^{1/2} \approx b_3 (\ln t), \quad (\sigma^6M_4(t))^{1/4} \approx b_4 (\ln  t).
    \label{eq:moments_comparison_KL}
\end{equation}
This approach is followed in Fig.~\ref{fig:KL_comparision_M3M4}(c) and Fig.~\ref{fig:KL_comparision_M3M4}(d) where we respectively plot  the dependence of $(\sigma^4 M_3(t))^{1/2}$ and $(\sigma^6M_4(t))^{1/4}$  on $\ln t$.
The obtained results clearly indicate linear growths of $(\sigma^4 M_3(t))^{1/2}$ and $(\sigma^6 M_4(t))^{1/4}$ with $\ln t$ for all displayed cases.
Similar results were also found for the remaining $6$ sets of $(W, m)$ parameters considered.
The slopes $b_3$ and $b_4$ are retrieved by fitting in all cases the numerical results with straight lines in the range $e^5 \leq t \leq e^9$.
These slopes are presented against $W$ in the insets of Fig.~\ref{fig:KL_comparision_M3M4}(c) and  Fig.~\ref{fig:KL_comparision_M3M4}(d), confirming the fact that their values are practically independent of the disorder strength $W$.

To sum up this section, we found that an initially localized wave-packet at the center of the disordered SSH chain exhibits AL (i.e. a halt in spreading for all times) within both the system's topologically trivial and non-trivial regions.
However, at the topological transition, this wave-packet is anomalously diffusing within the chain.
The time evolution of the wave-packet's moments follow logarithmic laws of time pass a critical order, while below that critical order, they asymptotically converge.

\subsection{\label{subsec:subsec:wave-packet_nonlinear}The effect of self-interaction potentials}
\begin{figure}
    \centering
    \includegraphics[width=\columnwidth]{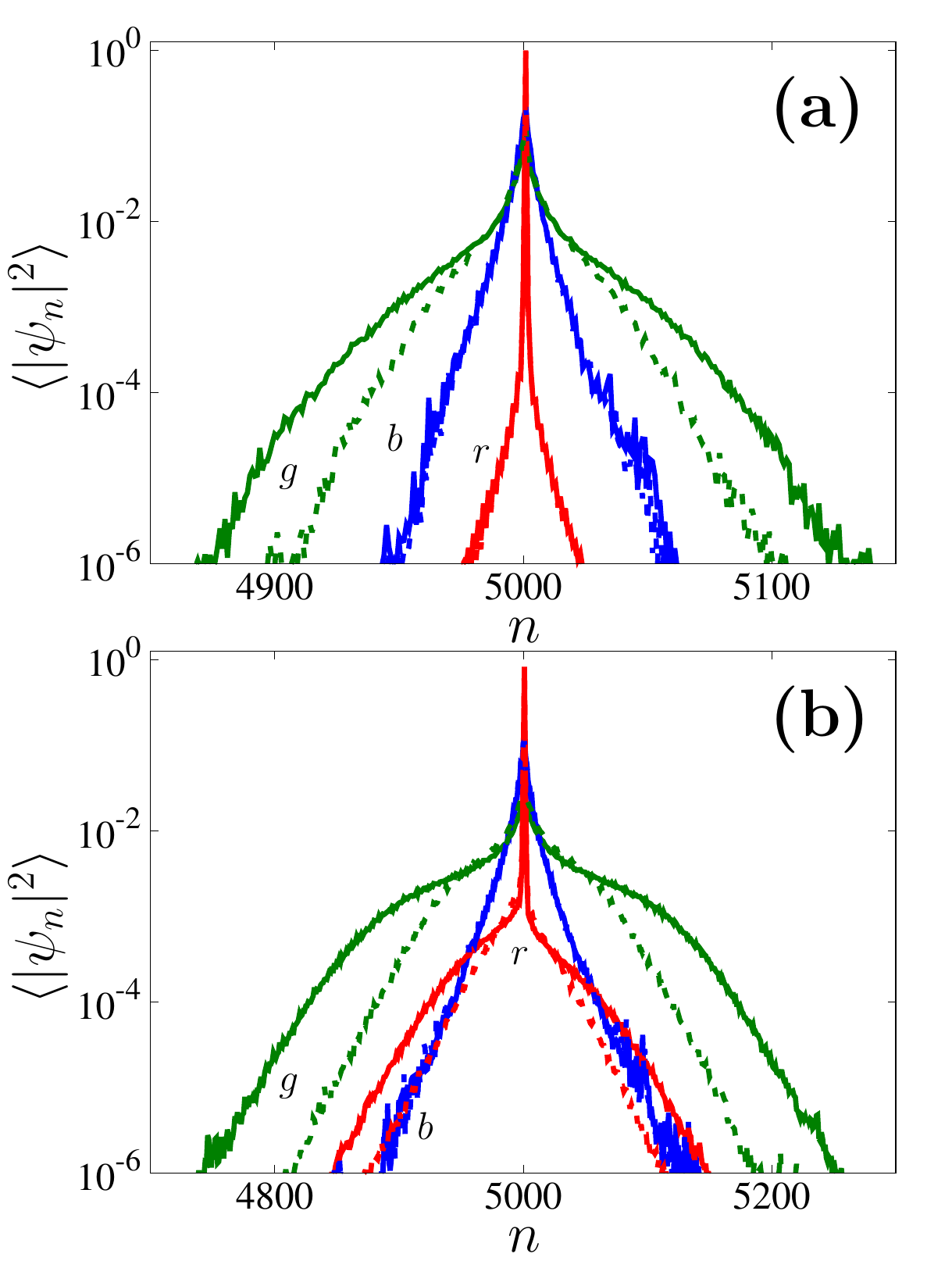}
    \caption{
        Average amplitude distribution $\langle \lvert \psi_n \rvert^2 \rangle$ at time $t\approx 10^5$ [dotted curves] and $t\approx 10^6$ [continuous curves] for the disordered nonlinear SSH chain [Eq.~\eqref{eq:eq_motion_nonlinear_general}], for (a) the case $(W=2.04, m=0.6)$ with $g=0$ [blue (b) curves] $g=3.0$ [green (g) curves] and $g=30$ [red (r) curves] and (b) the case  $(W=1.0, m=0.6)$ with $g=0$ [blue (b) curves], $g=1$ [green (g) curves] and $g=5$ [red (r) curves].
        The $\rvert \phi_n\rvert^2$ values are averaged over $2000$ configurations of disorder.
        The continuous and dotted blue (b) [red (r)] curves in panel (a) and the blue (b) curves in panel (b) are practically overlapping.
    }
    \label{fig:averaged_profile_wave_nonlinear_01}
\end{figure}

We now move to another important question and study what are the consequences of the presence of on-site nonlinearity on the dynamics of an initial single-site wave-packet excitation [Eq.~\eqref{eq:initial_wave_packet}].
In particular we consider the following set of equations of motion~\cite{MS2021,JD2022},
\begin{equation}
    i\frac{d\psi_n}{dt} = H_{n} \psi_{n+1} + H_{n-1} \psi_{n-1} + g\lvert \psi_n\rvert^2 \psi_n,
    \label{eq:eq_motion_nonlinear_general}
\end{equation}
where the term $g\lvert \psi_n \rvert^2 \psi_n$ describes the self interaction of the field $\psi_n$ with itself (self-interaction) at site with index $n$.
Note that this nonlinear model conserves the total energy $\mathcal{H}_4 = \mathcal{H}_2 + \sum_n g\lvert \psi_n \rvert^4/2$, and norm $ \mathcal{A} = \sum_n \lvert \psi_n \rvert^2$ of the system.
Further, in Eq.~\eqref{eq:eq_motion_nonlinear_general},  the real parameter $g>0$, is the nonlinear coefficient whose strength is classified into three regions: weak, moderate and strong~\cite{SKKF2009,ES2021}.
This generalization of the SSH model, or of any  topological lattice, with self-interaction (on-site) terms are relevant in experiments, and provide simple generic equations to describe the propagation of high amplitude wave-packets in chains of coupled nonlinear optical waveguides~\cite{ESMBA1998}, atomic BEC in optical traps~\cite{TS2001}, and synthetic lattices of atomic momentum states~\cite{MADMHG2018} to name a few.

We perform numerical simulations of the nonlinear model using the  initial conditions given by Eq.~(\ref{eq:initial_wave_packet}). Below, we vary the parameter $g$  in order to increase the strength of nonlinearity and monitor its effects.
Characteristic averaged (over $2000$ configurations of disorder) amplitude distributions at times $t\approx 10^5$ and $10^{6}$ (respectively the continuous and dotted curves) are shown in Fig.~\ref{fig:averaged_profile_wave_nonlinear_01}(a),
for a topologically non-trivial case with $(W=2.04, m=0.6)$.
The latter case leads to negative hopping coefficients between certain sites of the lattice for a chosen configuration of disorder~\cite{MADMHG2018}.
In this figure we have considered three different values of the nonlinear coefficient with 
$g=0$ (blue curves), $g=3$  (green curves) and  $g=30$ (red curves). In fact, the profile for $g=0$ is taken as instants from Fig.~\ref{fig:norm_W_against_time_W_constant}(a) of App.~\ref{app:sec:wave_packet_spreading} where we have shown that the systems remains localized due to AL. 
On the other hand, for a moderate nonlinearity with $g=3$, we find that the $\langle \lvert \psi_n \rvert^2 \rangle$ extent is broadening with time since the wave-packet profile is wider at $t\approx 10^6$ 
compared to the $t\approx 10^{5}$ case.
This spreading can only originate from the nonlinear mode-mode interactions~\cite{S1993,FKS2009} which are typically chaotic~\cite{SGF2013,SMS2018,MSS2020}.

Furthermore, for $g=30$ (red curves), the wave-packet retains a sharp and pointy profile at all time, with extent much smaller than  the one seen in the linear case for $g=0$.
This phenomenon is termed  selftrapping dynamical behavior~\cite{KKFA2008,SKKF2009}.
It stems from the fact that single-site excitations practically lead to discrete breathers, which in the limit of strong nonlinearity, essentially occupy a single-site, and have a negligible amplitude on neighboring sites.

We repeat the same procedure as above, but this time for the case $(W=1, m=0.6)$ in the topologically nontrivial phase, for which the hopping energies between sites are positively defined for all configurations of disorder.
We note that, this set of parameters are for example related to in experiments of coupled optical waveguides~\cite{ESMBA1998}.
The results of these simulations are presented in Fig.~\ref{fig:averaged_profile_wave_nonlinear_01}(b) where we focus our attention on the cases with $g=0$ (blue curves), $g=1$ (green curves) and $g=5$ (red curves) which corresponds to the linear, weakly and strongly nonlinear regimes in our system [Eq.~\eqref{eq:eq_motion_nonlinear_general}]. 
We clearly see for $g=0$ ($g=1$) the existence of AL (spreading) of the wave-packet when comparing its extents at times $t=10^{5}$ (dotted curves) and $t=10^{6}$ (continuous curves).
On the other hand, for $g=5$, we find that, although the wave-packet tends to spread, it retains a rather sharp and pointy shape around the initial site of excitation.


\begin{figure}
    \centering
    \includegraphics[width=\columnwidth]{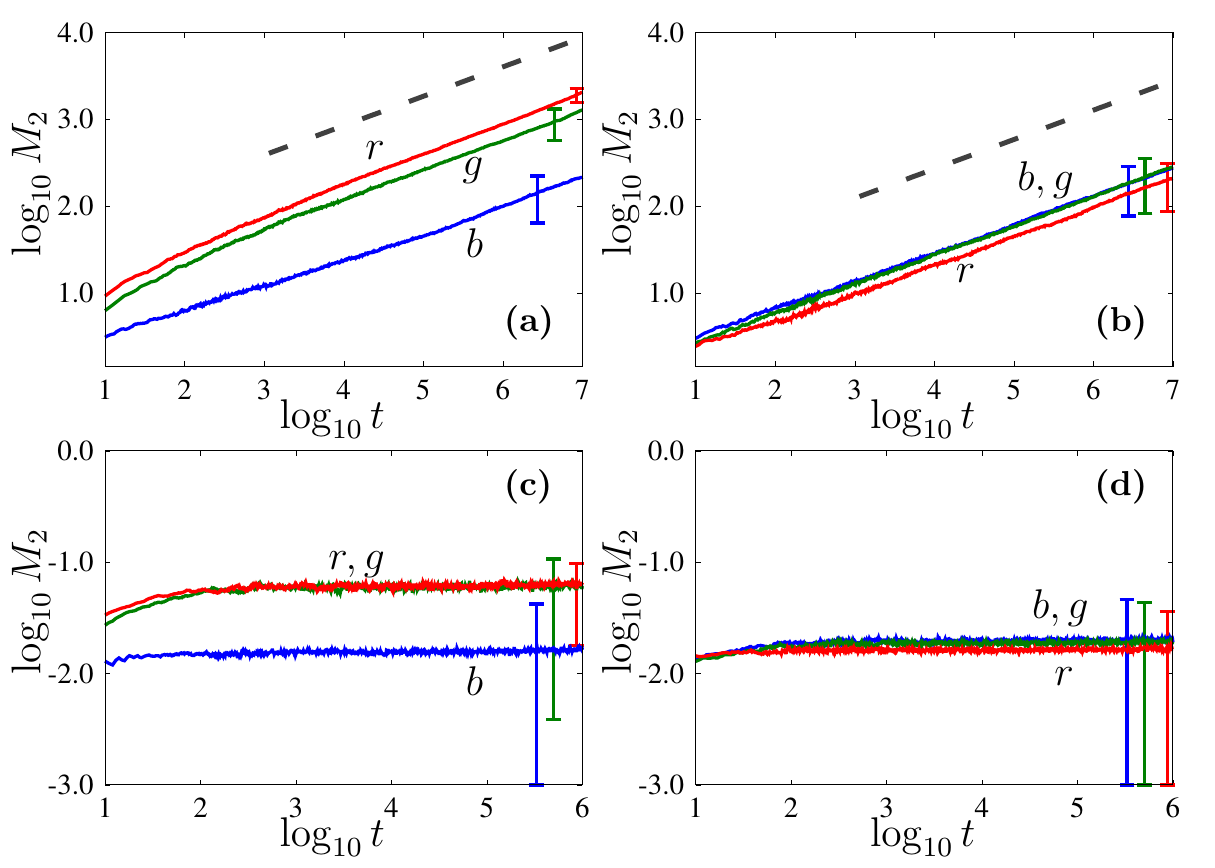}
    \caption{
        The time evolution of the second moment $M_2$ of the wave-packet for different $(W, m)$ parameter setups of the nonlinear system of Eq.~\eqref{eq:eq_motion_nonlinear_general} with (a)-(b) $g=3$ and (c)-(d) $g=30$. 
        Panels (a) and (c) correspond to the $(W=2.04, m=0.6)$, $(W=2.04, m=1.15)$ and $(W=2.04, m=1.6)$ cases [blue (b), green (g) and red (r) curves respectively] and panels (b) and (d) to the ones with $(W=3.0, m=0.6)$, $(W=4, m=0.6)$ and $(W=6, m=0.6)$ [blue (b), green (g) and red (r) curves respectively].
        The dashed lines in panels (a) and (b) guide the eye for slope $0.34$.
        The error bars of the same color as the curves, denote one standard deviation.
        Results are obtained by averaging over $100$ configurations of disorder.
    }
    \label{fig:wave-packet_second_moments_01}
\end{figure}

We completely characterize the spreading due to nonlinearity for $(W=2.04, m=0.6)$ with $g=3$, by following the time evolution of the second moment $M_2$.
Note that here, the results are averaged over $100$ configurations of disorder for which the time dependence of the moments shows a clear power law growth (synonym of chaotic nonlinear mode interactions) from the early stage of the evolution.
Using this procedure, we are able to probe the  effects of the nonlinear mode interactions on the long-time dynamics of localized wave-packets within the limited time period of our numerical simulations.
The obtained results are shown in Fig.~\ref{fig:wave-packet_second_moments_01} for the cases corresponding to  the square [Fig.~\ref{fig:wave-packet_second_moments_01}(a)] and circle [Fig.~\ref{fig:wave-packet_second_moments_01}(b)] points in the $(W, m)$ parameter space of Fig.~\ref{fig:topo_phase_diagram}(a).
The growth of the second moments $M_2$ are clearly visible at large times for all cases with moderate nonlinearity $g=3$. 
We fit these growths of the second moment with a power law $M_2 \propto t^{a_2}$ following 
\begin{equation}
    \log_{10} M_2 \approx a_2 \log_{10}t, 
    \label{eq:fitting_log} 
\end{equation}
in the interval $10^4\leq t \leq 10^7$.
The $a_2$ values obtained by this numerical fit  are $a_2 \approx 0.33$, $a_2 \approx 0.34$ and $a_2 \approx 0.35 $ respectively for the blue, green and red colored curves in Fig.~\ref{fig:wave-packet_second_moments_01}(a), and $a_2\approx 0.33$, $a_2 \approx 0.34$ and $a_2\approx 0.34$ for the cases of the blue, green and red colored curves in Fig.~\ref{fig:wave-packet_second_moments_01}(b).

It is worth noting that we have also checked the other wave-packet moments in our numerical simulations, although we do not report them  here to avoid repetition.
We found $M_1\propto t^{0.17}$, $M_3\propto t^{0.48}$ and $M_4\propto t^{0.61}$ for the cases in Fig.~\ref{fig:wave-packet_second_moments_01}(a) and
$M_1\propto t^{0.17}$, $M_3\propto t^{0.49}$ and $M_4\propto t^{0.64}$ for the ones in Fig.~\ref{fig:wave-packet_second_moments_01}(b).
These power exponents can be roughly related via a simple arithmetic sequence $a_q \approx q a_1$, where $a_1$ and $a_q$ are the power exponents of the first ($M_1$) and $q$th ($M_q$) moments respectively. 
Thus, this subdiffusive spreading is clearly different from the one observed at the topological transition of the linearized model discussed in Sec.~\ref{subsec:wave-packet_linear}.

We also performed a similar analysis for $g=30$ at $(W=2.04, m=0.6)$ in the system parameter space. 
In this context, we expect almost all the wave-packet to remain trapped around the position of the initial excitation for all cases, 
resulting to a saturation of the $M_2$, i.e. $M_2 \propto t^{0}$. 
In Figs.~\ref{fig:wave-packet_second_moments_01}(c)-(d), we find that this is indeed true as $M_2$ saturates for all studied cases to values $M_2 \approx 0$.
For the sake of completeness, we have also checked the other wave-packet moments $M_1$, $M_3$ and $M_4$ and found that these moments also saturate at large times to practically constant levels whose particular values depend on the parameters of the system.


These  numerically obtained values of the power law exponents are similar to the ones observed during the wave-packet subdiffusion in the context of the destruction of the AL by nonlinearity for models with diagonal disorder.
Indeed, expressing the wave-packet with respect to the system's NMs as  $\psi_n = \sum_\mu c_\mu B_{n, \mu}$ with $\psi_n$ and $c_\mu$ being the wave functions at site and mode with indices $n$ and $\mu$ respectively and $B_{n, \mu}$ being the $n\mbox{th}$ entry of the mode with index $\mu$, and substituting this expression into the equations of motion in real space [Eq.~\eqref{eq:eq_motion_nonlinear_general}], it follows that 
\begin{equation}
    i\frac{\partial c_\mu }{\partial t} = E_\mu c_\mu + g\sum_{\mu_1, \mu_2, \mu_3} V_{\mu, \mu_1, \mu_2, \mu_3} c_{\mu_1} c_{\mu_2}^\star c_{\mu_3}.
    \label{eq:eqs_mot_nonlinear_NMspace}
\end{equation}
In this equation $E_\mu$ is the eigenenergy of the $\mu$th mode ($\mu$ being the mode number) of the spectrum of the linearized system [Eq.~\eqref{eq:eigenvalue}], and
$
    V_{\mu, \mu_1, \mu_2, \mu_3} = \sum_{n} B_{n,\mu}^{\star} B_{n, \mu_1} B_{n, \mu_2}^\star B_{n, \mu_3}    
$
represents the overlap integral of four modes with $\left({}^{\star}\right)$ denoting the complex conjugate.
These equations of motion [Eq.~\eqref{eq:eqs_mot_nonlinear_NMspace}] were also obtained in a plethora of disordered systems in mechanics, optics and quantum physics (see e.g.~\cite{PS2008,MI2012,LIF2014,VFF2019,ES2021}).
In those works, it was found via analytical arguments and extensive numerical simulations that, for a single-site excitation, the moments $M_q$ of the wave-packet evolution asymptotically follow $M_q\propto t^{a_q}$ law, with $a_q = qa_1$, $a_1 = 1/6$.
These predictions are very close to our numerical results reported in this section.
The interesting result here is that this spreading power law is valid for both the situations of exponential localization of all NMs (AL) seen in the trivial and non-trivial topological phases, and in the presence of a large number of non-exponentially localized NMs (Dyson singularity) along the topological transition curve.
Furthermore, when the nonlinear strength is strong, following the {\it selftrapping theorem}~\cite{KKFA2008,SKKF2009}, little or no spreading of the wave-packet occurs across the entire $(W, m)$ parameter space.

\section{\label{sec:conclusion}Conclusions}
We studied in detail the long-time dynamics of a single-site excitation in the bulk of a disordered Su-Schrieffer-Heeger (SSH) chain in the linear and nonlinear regimes. 
A key feature of the disordered SSH model in its linear limit is the existence of a topologically non-trivial phase, separated to the trivial one by a curve in the parameter space of the hopping coefficient and the strength of the disorder.
Along that boundary curve, the localization length at small energies diverges in the thermodynamic limit. 
We performed direct numerical integrations of the system's equations of motion and computed the statistical moments of the amplitude distributions as  physical observables of the wave-packet dynamics.

In the linear regime,  our analysis showed, through variation of the parameters of the model, that inside the topological phases no propagation of the wave-packet is visible due to Anderson localization (AL).
On the other hand, at the topological transition an anomalous subdiffusion of the wave-packet is observed.
In order to quantify this spreading, we proposed an analytical expression of the wave-packet's moments, whose validity was verified through extensive numerical simulations.
We found that this anomalous spreading is characterized by a growth in time of the wave-packet's moments according to the law $M_q \propto \left(\ln t \right)^{2q-4}$, above a critical order $q=2$, for which we have $M_2 \propto \ln \left[\left(\ln t \right)^2 \right]$.
In addition, below that critical order, i.e. for $q<2$, the moments asymptotically saturate.

The anomalous subdiffusion along the topological transition is induced by the presence of a large number of non-exponentially localized modes at small energies, whose localization lengths grow with increasing lattice sizes.
This divergence of the localization length around the center of the frequency band is a signature of a topological transition in disordered systems supporting various topological phases.
Therefore, the wave-packet spreading can be used as an additional tool to identify topological transitions.
The former is extremely versatile, since it solely relies on the observation of the asymptotic dynamics of initially localized wave-packets of practically any shape and number of sites. 

We also studied the effect that the on-site nonlinearity has on the long-time dynamics of the wave-packet in the disordered SSH model.
To the best of our knowledge, this is the first time such results are reported.
In this context, we found that in the presence of weak and moderate strengths of nonlinearity, chaotic mode-mode interactions constitute the main mechanism of wave-packet dynamics.
This results to the same subdiffusive behavior of the wave-packet, irrespective of the modes' degree of localization.
Consequently, we numerically recovered in the whole system's parameter space the same power laws for time evolution of the wave-packet's moments, $M_q \propto t^{0.17q}$ for all orders, $q$.
More importantly, the power law obtained here is
very close to the one observed in models with on-site random scalar~\cite{FKS2009,GS2009,MI2012} and quasi-periodic~\cite{LLBDMF2012} potentials where $M_q \propto t^{q/6}$.

However at high strengths of nonlinearity we showed that the wave-packet spreading is partially or, even entirely,  suppressed due to the large shift of the nonlinear frequency of initially activated modes, which results to their failure to interact with their surroundings.
Consequently, nonlinear mode-mode interactions play an important role in the dynamics of nonlinear topological lattices and must be carefully scrutinized for the definition of reliable nonlinear topological markers.


We believe that the present work provides useful insights into the dynamics of disordered systems.
On one hand, the recent developments in topological physics have seen an upsurge on the investigation of the effects of symmetries in various physical phenomena. 
It is therefore natural to also look at what happens to the long-time dynamics of initially localized wave-packets in disordered systems when such symmetries are present in connection with the new topological characterization of matter.
On the other hand, understanding how the above pictures change when nonlinearity arises is also an interesting question.
As a final comment we note that although the present work focused on one dimensional lattices, it is rather natural to extend this analysis to systems of higher dimensions, in which we find richer topological properties. 
These aspects are currently under investigation and the related results will be reported in future publications.

\begin{acknowledgments}
The authors acknowledge the Centre for High Performance Computing (CHPC) of South Africa~\cite{CHPC2022} for providing computational resources that have contributed to the research results reported in this paper.
Ch.S. thanks Le Mans Universit\'e for its hospitality during his visit in 2022 where part of this work was carried out.
We also thank the three anonymous referees for their comments, which helped us improve the presentation
of our work.
\end{acknowledgments}

\appendix
\section{\label{app:sec:definitions}The chiral, projection, position and Hamiltonian matrices of the disordered SSH model}
Here we give the expressions of some important operators. 
Considering the SSH lattice of $L$ ($L$ is even) unit cells (see Fig.~\ref{fig:ssh_chain}), i.e.~with $N=2L$ sites, each unit cell is indexed with $l = [-L/2, -L/2+1, -L/2+2, \ldots, -L/2+L/2-2, -L/2+L/2-1, 0, 1, 2, \ldots, L/2-2, L/2-1]$.
Consequently the chiral ($\Gamma$), the projections into the $A$  ($\Gamma _A$) and $B$  ($\Gamma_B$) sublattices and the position ($\mathbb{X}$) operators respectively have the following forms~\cite{MADMHG2018,SKCATY2021}
\begin{widetext}
\begin{equation}
    \Gamma = 
    \begin{pmatrix}
        1 & 0 & 0 & 0 & \ldots \\
        0 & -1 & 0 & 0 & \ldots \\ 
        0 & 0 & 1 & 0 & \ldots \\ 
        0 & 0 & 0 & -1 & \ldots \\ 
        \vdots & \vdots & \vdots & \vdots & \ddots \\
        \end{pmatrix}_{N\times N}, \quad 
    \Gamma_A = 
    \begin{pmatrix}
        1 & 0 & 0 & 0 & \ldots \\
        0 & 0 & 0 & 0 & \ldots \\ 
        0 & 0 & 1 & 0 & \ldots \\ 
        0 & 0 & 0 & 0 & \ldots \\ 
        \vdots & \vdots & \vdots & \vdots & \ddots \\
        \end{pmatrix}_{N\times N}, \quad
    \Gamma_B = 
    \begin{pmatrix}
        0 & 0 & 0 & 0 & \ldots \\
        0 & 1 & 0 & 0 & \ldots \\ 
        0 & 0 & 0 & 0 & \ldots \\ 
        0 & 0 & 0 & 1 & \ldots \\ 
        \vdots & \vdots & \vdots & \vdots & \ddots \\
    \end{pmatrix}_{N\times N}
    \label{eq:chiral_matrices_01}
\end{equation}
\begin{equation}
    \mathbb{X} = 
    \begin{pmatrix}
        -\frac{L}{2} & 0 & 0 & 0 & \ldots & 0 & 0 \\ 
        0 & -\frac{L}{2} & 0 & 0 & \ldots & 0 & 0 \\ 
        0 & 0 & -\frac{L}{2}+1 & 0 & \ldots & 0 & 0 & \\ 
        0 & 0 & 0 & -\frac{L}{2}+1 & \ldots & 0 & 0 \\ 
        \vdots & \vdots & \vdots & \vdots & \ddots & \vdots & \vdots \\ 
        0 & 0 & 0 & 0 & \ldots & \frac{L}{2}-1 & 0 \\ 
        0 & 0 & 0 & 0 & \ldots & 0 & \frac{L}{2}-1 \\
    \end{pmatrix}_{N\times N}.
    \label{eq:position_matrix}
\end{equation}
\end{widetext}
Moreover, the expression of the random Hamiltonian matrix $\mathbb{H}$ [Eq.~\eqref{eq:eigenvalue}] of the disordered SSH model in its linearized limit reads
\begin{widetext}
\begin{equation}
    \mathbb{H} = 
    \begin{pmatrix}
         0 & m + W \epsilon_1 & 0 & 0 & \ldots & 0 & 0 \\ 
        m + W \epsilon_1 & 0 & 1 + \frac{W}{2} \epsilon_2 & 0 & \ldots & 0 & 0 \\
        0 & 1 + \frac{W}{2} \epsilon_2 & 0 & m + W \epsilon_3 & \ldots & 0 & 0 \\ 
        0 & 0 & m + W \epsilon_3 & 0 & \ldots & 0 & 0 \\ 
        \vdots & \vdots & \vdots & \vdots & \ddots & \vdots & \vdots \\ 
        0 & 0 & 0 & 0 & \ldots & 0 & m + W \epsilon_{N-1} \\ 
        0 & 0 & 0 & 0  & \ldots & m+ W \epsilon_{N-1} & 0 
    \end{pmatrix}_{N\times N},
    \label{eq:Hamiltonian_matrix}
\end{equation}
\end{widetext}
where $m$ and $1$ are the intra- and inter-cell hoppings of the periodic SSH chain, $\{ \epsilon_i\}_{i=1,N}$ characterizes random parameters drawn on the interval $[-1/2, 1/2]$ and $W$ is the strength of disorder.

\section{\label{app:sec:zero_energy}``Zero" energy states in the disordered SSH chain}

\begin{figure}
    \centering
    \includegraphics[width=0.85\columnwidth]{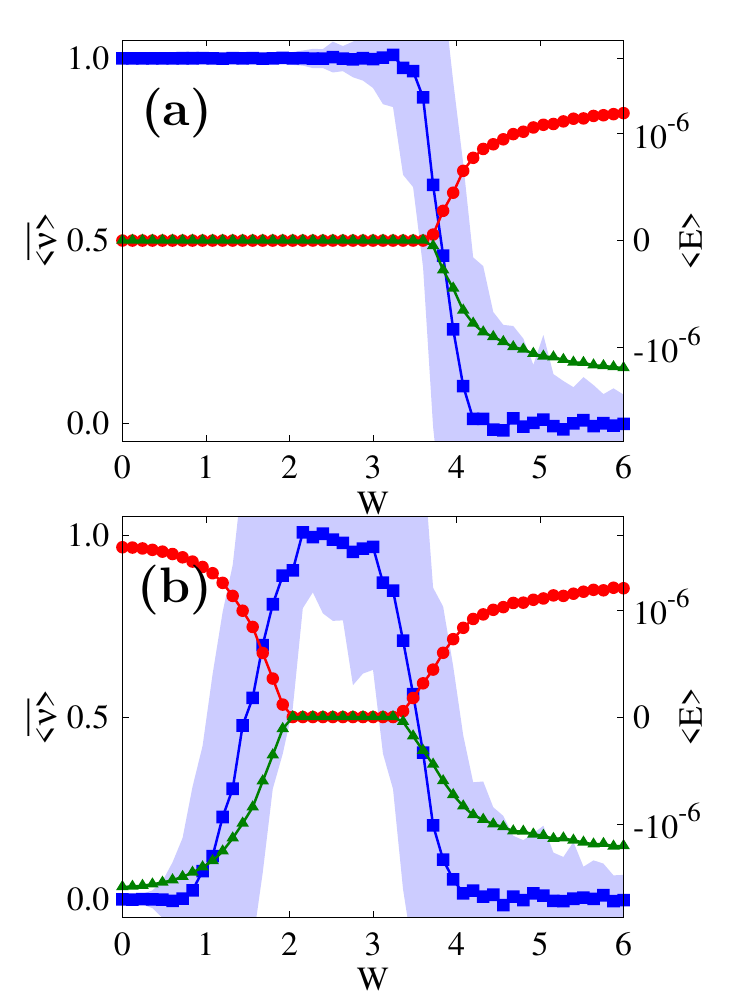}
    \caption{
    Sections of the topological phase diagram at (a) $(W, m=0.6)$ [dashed-dotted-dotted red line in Fig.~\ref{fig:topo_phase_diagram} (a)] and (b) $(W, m=1.05)$ [dashed-dotted green line in Fig.~\ref{fig:topo_phase_diagram} (a)].
    On top of the winding number, $\langle \overline{\nu}\rangle$ (blue line-connected squares) we superimpose the disorder averaged energies of the modes closest to the center of the spectrum from the negative (green line-connected triangles) and positive (red line-connected circles) sides of the origin.
    The shaded areas around the blue dotted curves indicate one standard deviation.
    }
    \label{fig:minuslogE}
\end{figure}

Here, we study in detail the properties of the two closest to zero energy states along different regions in the $(W,m)$ parameter space.
In Fig.~\ref{fig:minuslogE}(a), we plot the winding number, $\langle \overline{\nu} \rangle$ [blue line-connected squares] along the line $m=0.6$ [dashed-dotted-dotted red line in Fig.~\ref{fig:topo_phase_diagram}(a)]. 
At around $W\approx 3.75$, a topological transition appears.
In Fig.~\ref{fig:minuslogE}(a), we superimpose the average over $10^3$ configurations of disorder of the two energies $\langle E\rangle$ closest to zero. 
We clearly see that for a non-zero winding number two modes persist at zero energy. On the other hand, when the winding goes to zero these two modes clearly acquire finite energy values.

In Fig.~\ref{fig:minuslogE}(b), we observe a similar behavior when we follow the parametric line $m=1.05$ [dashed-dotted green line in Fig.~\ref{fig:topo_phase_diagram}(a)].
The main difference here compared to the case of Fig.~\ref{fig:minuslogE}(a) is the emergence of a non-trivial topological phase from a trivial one, varying the strength of disorder in the range $2\lesssim W \lesssim 3.5$.
This so called disorder-induced topological phase, is also accompanied by a pair of zero energy states.

\begin{figure}
    \centering
    \includegraphics[width=\columnwidth]{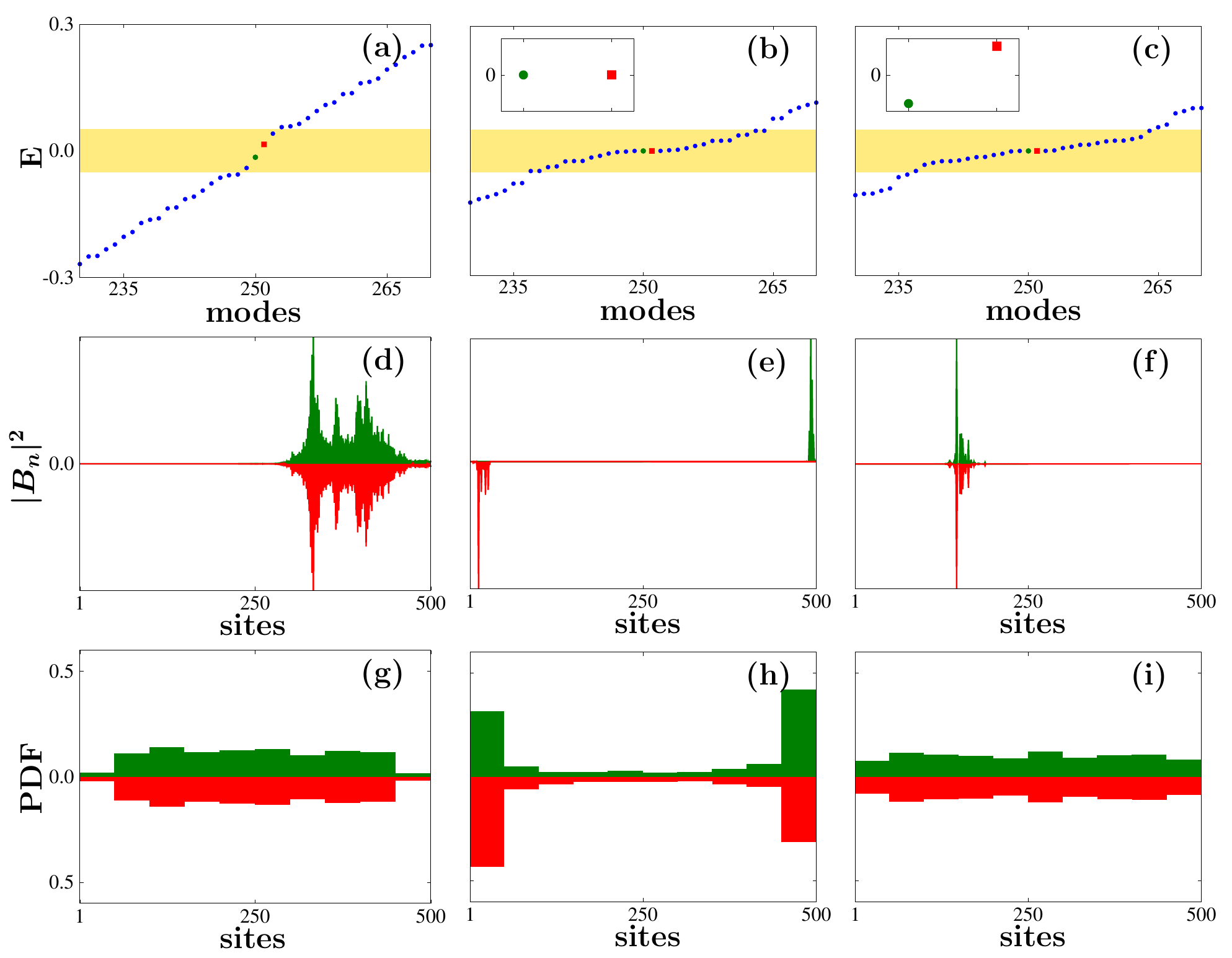}
    \caption{
         Results for representative configurations of disorder for three parameter sets along the line $m=1.05$ in the $(W, m)$ space with $W=0.5$ [(a), (d), (g)], $W=2.98$ [(b), (e), (h)] and $W=4.95$ [(c), (f), (i)].
         [(a)-(c)] The energy spectra, $E$ as function of the mode number. 
        The inset in (b)-(c) are zooms of the spectra around the center within the range $E\in [-10^{-4}, 10^{-4}]$.
        The yellow strips indicate the width of the energy gap for the clean chain ($W=0$). 
         [(d)-(f)] Amplitude $\lvert B _n\rvert^2$ of the modes with the smallest energies for the same configurations of disorder as in panels (a)-(c).
         [(g)-(i)] PDFs of the center of mass of the modes closest to the center of the spectra, generated within intervals of size $50$ sites along the chain.
        In all panels, the green (red) color, corresponds to the mode closest to the center of the energy spectrum with $E<0$ ($E>0$).
    }
    \label{fig:COM_01a}
\end{figure}

 The question now is where the zero, or the closest to zero, energy states in both the non-trivial and trivial topological phases, are located within the chain.
In order to address this question, we consider three sets of parameters for $m=1.05$ [i.e.~along the dashed-dotted green line in Fig.~\ref{fig:topo_phase_diagram}(a)] with $W=0.5$ [Figs.~\ref{fig:COM_01a}(a),(d),(g)], $W=2.98$ [Figs.~\ref{fig:COM_01a}(b),(e),(h)] and $W=4.5$ [Figs.~\ref{fig:COM_01a}(c),(f),(i)].
The first and last cases are located in the topologically trivial phase [Fig.~\ref{fig:topo_phase_diagram}(a)], while the middle one corresponds to the non-trivial counterpart.  
We first select representative configurations of disorder for these  parameter sets considering lattices with $N=500$ particles, and plot their energy bands in Figs.~\ref{fig:COM_01a}(a)-(c).
From the results of these figures we confirm that there are zero energy modes only in the case of $(W=2.98, m=1.05)$, inside the topologically non-trivial phase, as shown in the inset of Fig.~\ref{fig:COM_01a}(b), while no such modes exist in the cases with $(W=0.5, m=1.05)$ and $(W=4.95, m=1.05)$ which are within the trivial phase of the chain.
In addition, we plot in Figs.~\ref{fig:COM_01a}(d)-(f) the amplitudes of the modes with the smallest energies and show that they are always localized.
Nonetheless, in the topologically trivial cases [(Fig.~\ref{fig:COM_01a}(d) and Fig.~\ref{fig:COM_01a}(f)], we see that these modes are located inside the bulk of the lattice  while for the non-trivial counterpart [Fig.~\ref{fig:COM_01a}(e)], we observe that the modes closest to the center of the energy band are located at the boundaries of the chain.

In order to support these rather qualitative results, we perform a statistical analysis on the states with energy closest to the origin, by computing the probability distribution function (PDF) of their center of mass 
\begin{equation}
    \overline{n}_\mu = \sum_{n=1}^{N} n \lvert B_{ n,\mu}\rvert^2, 
    \label{eq:COM}
\end{equation}
where $n$ and $\mu$ being the site and mode indices, with $\sum_{n=1}^{N} \lvert B_{n,\mu} \rvert^2 = 1$.
Thus, as in the mechanical analogue of the disordered SSH chain~\cite{SKCATY2021}, the numerically calculated PDFs of the center of mass for about $10^3$ configurations of disorder, reveal that the modes with the smallest energies in the topologically trivial phase can be located at any site along the chain with practically equal likelihood [Fig.~\ref{fig:COM_01a}(g) and Fig.~\ref{fig:COM_01a}(i)].
On the other hand, within the non-trivial topological phase, these modes are most likely to be situated at the edges of the chain [Fig.~\ref{fig:COM_01a}(h)].

Let us discuss a bit more the trend of the PDFs depicted in Fig.~\ref{fig:COM_01a}(g) in which we observed practically zero values at the edge of the chain. 
Such behavior comes from the fact that at small $W$, the modes are rather extended [see e.g. Fig.~\ref{fig:COM_01a}(d)]. 
Consequently, the computation of the center of mass [Eq.~\eqref{eq:COM}] of the modes located at, or close to, the boundaries of the lattice, results to a value inside the bulk of the chain. 

\section{\label{app:sec:wave_packet_spreading}Wave-packet dynamics in the disordered SSH chain}
\begin{figure*}
    \centering
    \includegraphics[width=\textwidth]{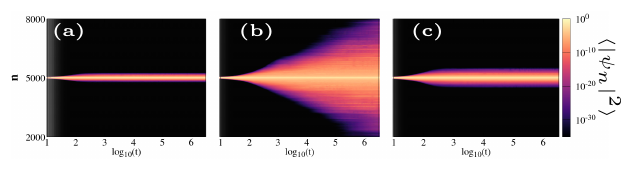}
    \caption{Similar to Fig.~\ref{fig:norm_W_against_time_m_constant}, but for (a) $m=0.6$, (b) $m=1.15$ and (c) $m=1.7$ with $W=2.04$ line corresponding to the blue, green and red squares in Fig.~\ref{fig:topo_phase_diagram}(a).
       The amplitudes are averaged over $2000$ configurations of disorder.
}
    \label{fig:norm_W_against_time_W_constant}
\end{figure*}

We provide additional numerical simulations to further substantiate the observations on the dynamics of localized wave-packet excitations in the disordered SSH chain (Fig.~\ref{fig:ssh_chain}). 
In Fig.~\ref{fig:norm_W_against_time_W_constant}, we plot the time evolution of the wave-packet averaged over  $2000$ configurations of disorder, for three representative parameter sets along the $W=2.04$ line with $m=0.6$ inside the topologically non-trivial region [blue square in Fig.~\ref{fig:topo_phase_diagram}(a)], $m=1.15$ at the topological transition curve [green square in Fig.~\ref{fig:topo_phase_diagram}(a)] and $m=1.7$ within the topologically trivial phase [red square in Fig.~\ref{fig:topo_phase_diagram}(a)].
We find that within the two topologically distinct regions, a halt in the spreading of the wave-packet after a finite time occurs [Figs.~\ref{fig:norm_W_against_time_W_constant}(a) and~\ref{fig:norm_W_against_time_W_constant}(c)]. 
However for the case at the topological transition curve, a continuous wave-packet growth is seen up to the largest time of our simulations ($t\approx 10^6$).

\begin{figure}
    \centering
    \includegraphics[width=\columnwidth]{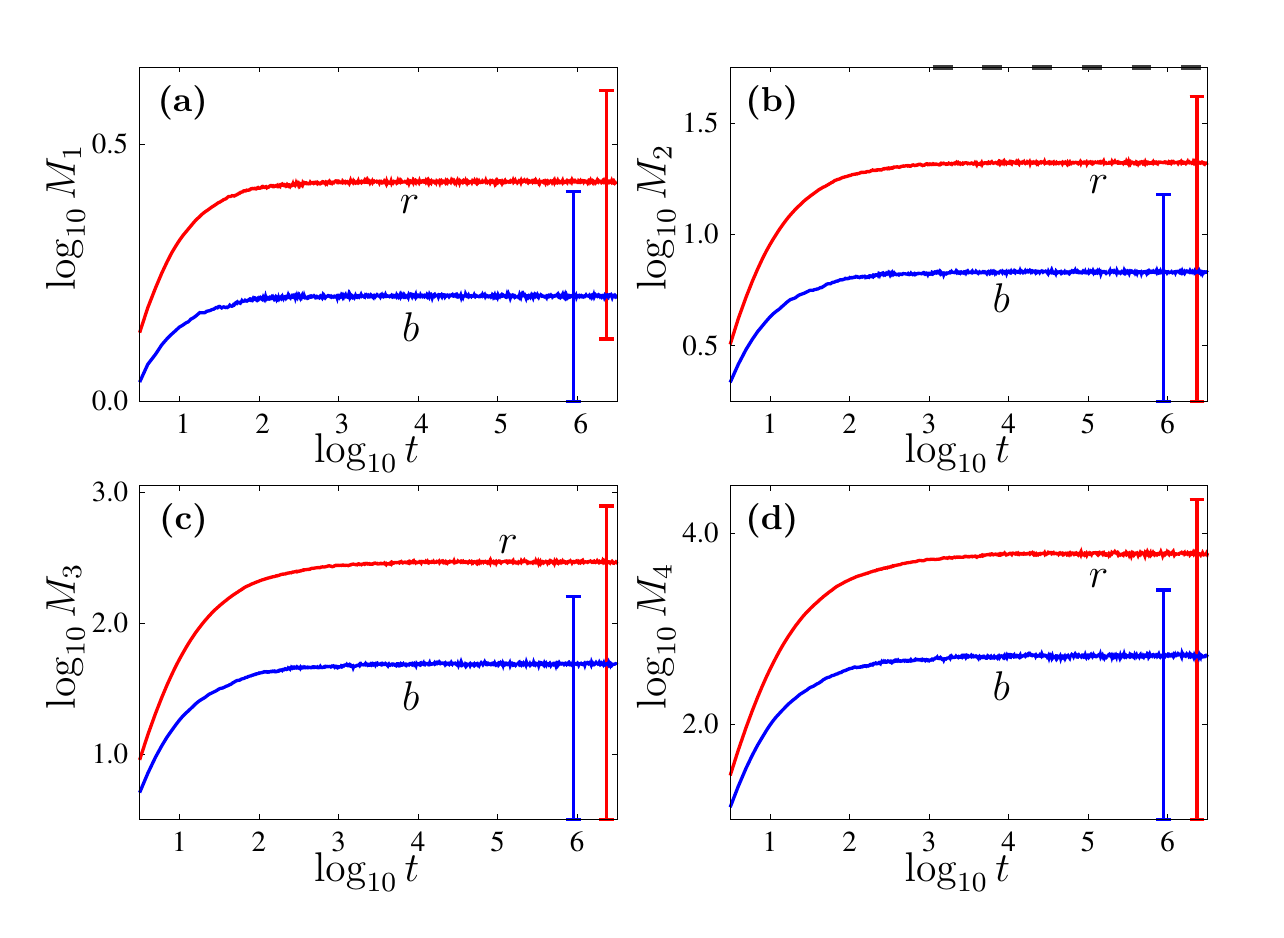}
    \caption{
        Similar to Fig.~\ref{fig:Mq_m_constant} but for $(W=2.04, m=0.6)$ [blue (b) curves] and $(W=2.04, m=1.17)$ [red (r) curves].
        The results were obtained by averaging the moments over $9000$ configurations of disorder.
    }
    \label{fig:Mq_W_constant}
\end{figure}

Figure~\ref{fig:Mq_W_constant} present the time evolution of the moments $M_q$, $q=1,2,3,4$ of the wave-packet averaged over $9000$ configurations of disorder, for the two representatives sets of parameters 
of Fig.~\ref{fig:norm_W_against_time_W_constant}(a) with $m=0.6$ (blue curves in Fig.~\ref{fig:Mq_W_constant}) and Fig.~\ref{fig:norm_W_against_time_W_constant}(c) with $m=1.7$ (red curves in Fig.~\ref{fig:Mq_W_constant}).
For these cases, the $M_1(t)$ [Fig.~\ref{fig:Mq_W_constant}(a)], $M_2(t)$ [Fig.~\ref{fig:Mq_W_constant}(b)], $M_3(t)$ [Fig.~\ref{fig:Mq_W_constant}(c)] and $M_4(t)$ [Fig.~\ref{fig:Mq_W_constant}(d)] tend to grow at the early stage of the evolution.
At larger times, it is clear that all moments asymptotically saturate to different values depending on the $(W, m)$ point.
It is worth emphasizing that the outcomes of the same analysis for the case of Fig.~\ref{fig:norm_W_against_time_W_constant}(b) are presented in Fig.~\ref{fig:KL_comparision_M3M4}. 
In this case a clear growth of the moments is observed, in agreement with the wave-packet spreading observed in Fig.~\ref{fig:norm_W_against_time_W_constant}(b).

\let\itshape\upshape
\normalem
\bibliography{References}

\providecommand{\noopsort}[1]{}\providecommand{\singleletter}[1]{#1}%
\begin{thebibliography}{81}%
\makeatletter
\providecommand \@ifxundefined [1]{%
 \@ifx{#1\undefined}
}%
\providecommand \@ifnum [1]{%
 \ifnum #1\expandafter \@firstoftwo
 \else \expandafter \@secondoftwo
 \fi
}%
\providecommand \@ifx [1]{%
 \ifx #1\expandafter \@firstoftwo
 \else \expandafter \@secondoftwo
 \fi
}%
\providecommand \natexlab [1]{#1}%
\providecommand \enquote  [1]{``#1''}%
\providecommand \bibnamefont  [1]{#1}%
\providecommand \bibfnamefont [1]{#1}%
\providecommand \citenamefont [1]{#1}%
\providecommand \href@noop [0]{\@secondoftwo}%
\providecommand \href [0]{\begingroup \@sanitize@url \@href}%
\providecommand \@href[1]{\@@startlink{#1}\@@href}%
\providecommand \@@href[1]{\endgroup#1\@@endlink}%
\providecommand \@sanitize@url [0]{\catcode `\\12\catcode `\$12\catcode
  `\&12\catcode `\#12\catcode `\^12\catcode `\_12\catcode `\%12\relax}%
\providecommand \@@startlink[1]{}%
\providecommand \@@endlink[0]{}%
\providecommand \url  [0]{\begingroup\@sanitize@url \@url }%
\providecommand \@url [1]{\endgroup\@href {#1}{\urlprefix }}%
\providecommand \urlprefix  [0]{URL }%
\providecommand \Eprint [0]{\href }%
\providecommand \doibase [0]{http://dx.doi.org/}%
\providecommand \selectlanguage [0]{\@gobble}%
\providecommand \bibinfo  [0]{\@secondoftwo}%
\providecommand \bibfield  [0]{\@secondoftwo}%
\providecommand \translation [1]{[#1]}%
\providecommand \BibitemOpen [0]{}%
\providecommand \bibitemStop [0]{}%
\providecommand \bibitemNoStop [0]{.\EOS\space}%
\providecommand \EOS [0]{\spacefactor3000\relax}%
\providecommand \BibitemShut  [1]{\csname bibitem#1\endcsname}%
\let\auto@bib@innerbib\@empty
\bibitem [{\citenamefont {Lu}\ \emph {et~al.}(2014)\citenamefont {Lu},
  \citenamefont {Joannopoulos},\ and\ \citenamefont
  {Solja{\v{c}}i{\'c}}}]{LJS2014}%
  \BibitemOpen
  \bibfield  {author} {\bibinfo {author} {\bibfnamefont {L.}~\bibnamefont
  {Lu}}, \bibinfo {author} {\bibfnamefont {J.~D.}\ \bibnamefont
  {Joannopoulos}}, \ and\ \bibinfo {author} {\bibfnamefont {M.}~\bibnamefont
  {Solja{\v{c}}i{\'c}}},\ }\bibfield  {title} {\enquote {\bibinfo {title}
  {Topological photonics},}\ }\href {\doibase 10.1038/nphoton.2014.248}
  {\bibfield  {journal} {\bibinfo  {journal} {Nat. Photonics}\ }\textbf
  {\bibinfo {volume} {8}},\ \bibinfo {pages} {821} (\bibinfo {year}
  {2014})}\BibitemShut {NoStop}%
\bibitem [{\citenamefont {Ozawa}\ \emph {et~al.}(2019)\citenamefont {Ozawa},
  \citenamefont {Price}, \citenamefont {Amo}, \citenamefont {Goldman},
  \citenamefont {Hafezi}, \citenamefont {Lu}, \citenamefont {Rechtsman},
  \citenamefont {Schuster}, \citenamefont {Simon}, \citenamefont {Zilberberg},\
  and\ \citenamefont {Carusotto}}]{OPAGHLRSSZC2019}%
  \BibitemOpen
  \bibfield  {author} {\bibinfo {author} {\bibfnamefont {T.}~\bibnamefont
  {Ozawa}}, \bibinfo {author} {\bibfnamefont {H.~M.}\ \bibnamefont {Price}},
  \bibinfo {author} {\bibfnamefont {A.}~\bibnamefont {Amo}}, \bibinfo {author}
  {\bibfnamefont {N.}~\bibnamefont {Goldman}}, \bibinfo {author} {\bibfnamefont
  {M.}~\bibnamefont {Hafezi}}, \bibinfo {author} {\bibfnamefont
  {L.}~\bibnamefont {Lu}}, \bibinfo {author} {\bibfnamefont {M.~C.}\
  \bibnamefont {Rechtsman}}, \bibinfo {author} {\bibfnamefont {D.}~\bibnamefont
  {Schuster}}, \bibinfo {author} {\bibfnamefont {J.}~\bibnamefont {Simon}},
  \bibinfo {author} {\bibfnamefont {O.}~\bibnamefont {Zilberberg}}, \ and\
  \bibinfo {author} {\bibfnamefont {I.}~\bibnamefont {Carusotto}},\ }\bibfield
  {title} {\enquote {\bibinfo {title} {Topological photonics},}\ }\href
  {\doibase 10.1103/RevModPhys.91.015006} {\bibfield  {journal} {\bibinfo
  {journal} {Rev. Mod. Phys.}\ }\textbf {\bibinfo {volume} {91}},\ \bibinfo
  {pages} {015006} (\bibinfo {year} {2019})}\BibitemShut {NoStop}%
\bibitem [{\citenamefont {Segev}\ and\ \citenamefont {Bandres}(2021)}]{SB2021}%
  \BibitemOpen
  \bibfield  {author} {\bibinfo {author} {\bibfnamefont {M.}~\bibnamefont
  {Segev}}\ and\ \bibinfo {author} {\bibfnamefont {M.~A.}\ \bibnamefont
  {Bandres}},\ }\bibfield  {title} {\enquote {\bibinfo {title} {Topological
  photonics: {W}here do we go from here?}}\ }\href {\doibase
  doi:10.1515/nanoph-2020-0441} {\bibfield  {journal} {\bibinfo  {journal}
  {Nanophotonics}\ }\textbf {\bibinfo {volume} {10}},\ \bibinfo {pages} {425}
  (\bibinfo {year} {2021})}\BibitemShut {NoStop}%
\bibitem [{\citenamefont {Hasan}\ and\ \citenamefont {Kane}(2010)}]{HK2010}%
  \BibitemOpen
  \bibfield  {author} {\bibinfo {author} {\bibfnamefont {M.~Z.}\ \bibnamefont
  {Hasan}}\ and\ \bibinfo {author} {\bibfnamefont {C.~L.}\ \bibnamefont
  {Kane}},\ }\bibfield  {title} {\enquote {\bibinfo {title} {Colloquium:
  Topological insulators},}\ }\href {\doibase 10.1103/RevModPhys.82.3045}
  {\bibfield  {journal} {\bibinfo  {journal} {Rev. Mod. Phys.}\ }\textbf
  {\bibinfo {volume} {82}},\ \bibinfo {pages} {3045} (\bibinfo {year}
  {2010})}\BibitemShut {NoStop}%
\bibitem [{\citenamefont {Qi}\ and\ \citenamefont {Zhang}(2011)}]{QZ2011}%
  \BibitemOpen
  \bibfield  {author} {\bibinfo {author} {\bibfnamefont {X.-L.}\ \bibnamefont
  {Qi}}\ and\ \bibinfo {author} {\bibfnamefont {S.-C.}\ \bibnamefont {Zhang}},\
  }\bibfield  {title} {\enquote {\bibinfo {title} {Topological insulators and
  superconductors},}\ }\href {\doibase 10.1103/RevModPhys.83.1057} {\bibfield
  {journal} {\bibinfo  {journal} {Rev. Mod. Phys.}\ }\textbf {\bibinfo {volume}
  {83}},\ \bibinfo {pages} {1057} (\bibinfo {year} {2011})}\BibitemShut
  {NoStop}%
\bibitem [{\citenamefont {Ezawa}(2018)}]{E2018a}%
  \BibitemOpen
  \bibfield  {author} {\bibinfo {author} {\bibfnamefont {M.}~\bibnamefont
  {Ezawa}},\ }\bibfield  {title} {\enquote {\bibinfo {title} {Higher-order
  topological electric circuits and topological corner resonance on the
  breathing {K}agome and pyrochlore lattices},}\ }\href {\doibase
  10.1103/PhysRevB.98.201402} {\bibfield  {journal} {\bibinfo  {journal} {Phys.
  Rev. B}\ }\textbf {\bibinfo {volume} {98}},\ \bibinfo {pages} {201402}
  (\bibinfo {year} {2018})}\BibitemShut {NoStop}%
\bibitem [{\citenamefont {Smirnova}\ \emph {et~al.}(2020)\citenamefont
  {Smirnova}, \citenamefont {Leykam}, \citenamefont {Chong},\ and\
  \citenamefont {Kivshar}}]{SLCK2020}%
  \BibitemOpen
  \bibfield  {author} {\bibinfo {author} {\bibfnamefont {D.}~\bibnamefont
  {Smirnova}}, \bibinfo {author} {\bibfnamefont {D.}~\bibnamefont {Leykam}},
  \bibinfo {author} {\bibfnamefont {Y.}~\bibnamefont {Chong}}, \ and\ \bibinfo
  {author} {\bibfnamefont {Y.}~\bibnamefont {Kivshar}},\ }\bibfield  {title}
  {\enquote {\bibinfo {title} {Nonlinear topological photonics},}\ }\href
  {\doibase 10.1063/1.5142397} {\bibfield  {journal} {\bibinfo  {journal}
  {Appl. Phys. Rev.}\ }\textbf {\bibinfo {volume} {7}},\ \bibinfo {pages}
  {021306} (\bibinfo {year} {2020})}\BibitemShut {NoStop}%
\bibitem [{\citenamefont {Huber}(2016)}]{H2016}%
  \BibitemOpen
  \bibfield  {author} {\bibinfo {author} {\bibfnamefont {S.~D.}\ \bibnamefont
  {Huber}},\ }\bibfield  {title} {\enquote {\bibinfo {title} {Topological
  mechanics},}\ }\href {\doibase 10.1038/nphys3801} {\bibfield  {journal}
  {\bibinfo  {journal} {Nat. Phys.}\ }\textbf {\bibinfo {volume} {12}},\
  \bibinfo {pages} {621} (\bibinfo {year} {2016})}\BibitemShut {NoStop}%
\bibitem [{\citenamefont {Süsstrunk}\ and\ \citenamefont
  {Huber}(2016)}]{SH2016}%
  \BibitemOpen
  \bibfield  {author} {\bibinfo {author} {\bibfnamefont {R.}~\bibnamefont
  {Süsstrunk}}\ and\ \bibinfo {author} {\bibfnamefont {S.~D.}\ \bibnamefont
  {Huber}},\ }\bibfield  {title} {\enquote {\bibinfo {title} {Classification of
  topological phonons in linear mechanical metamaterials},}\ }\href {\doibase
  10.1073/pnas.1605462113} {\bibfield  {journal} {\bibinfo  {journal} {Proc.
  Natl. Acad. Sci. U.S.A.}\ }\textbf {\bibinfo {volume} {113}},\ \bibinfo
  {pages} {E4767} (\bibinfo {year} {2016})}\BibitemShut {NoStop}%
\bibitem [{\citenamefont {Ma}\ \emph {et~al.}(2019)\citenamefont {Ma},
  \citenamefont {Xiao},\ and\ \citenamefont {Chan}}]{MXC2019}%
  \BibitemOpen
  \bibfield  {author} {\bibinfo {author} {\bibfnamefont {G.}~\bibnamefont
  {Ma}}, \bibinfo {author} {\bibfnamefont {M.}~\bibnamefont {Xiao}}, \ and\
  \bibinfo {author} {\bibfnamefont {C.~T.}\ \bibnamefont {Chan}},\ }\bibfield
  {title} {\enquote {\bibinfo {title} {Topological phases in acoustic and
  mechanical systems},}\ }\href {\doibase 10.1038/s42254-019-0030-x} {\bibfield
   {journal} {\bibinfo  {journal} {Nat. Rev. Phys.}\ }\textbf {\bibinfo
  {volume} {1}},\ \bibinfo {pages} {281} (\bibinfo {year} {2019})}\BibitemShut
  {NoStop}%
\bibitem [{\citenamefont {Xin}\ \emph {et~al.}(2020)\citenamefont {Xin},
  \citenamefont {Siyuan}, \citenamefont {Harry}, \citenamefont {Minghui},\ and\
  \citenamefont {Yanfeng}}]{XSHMY2020}%
  \BibitemOpen
  \bibfield  {author} {\bibinfo {author} {\bibfnamefont {L.}~\bibnamefont
  {Xin}}, \bibinfo {author} {\bibfnamefont {Y.}~\bibnamefont {Siyuan}},
  \bibinfo {author} {\bibfnamefont {L.}~\bibnamefont {Harry}}, \bibinfo
  {author} {\bibfnamefont {L.}~\bibnamefont {Minghui}}, \ and\ \bibinfo
  {author} {\bibfnamefont {C.}~\bibnamefont {Yanfeng}},\ }\bibfield  {title}
  {\enquote {\bibinfo {title} {Topological mechanical metamaterials: {A} brief
  review},}\ }\href {\doibase https://doi.org/10.1016/j.cossms.2020.100853}
  {\bibfield  {journal} {\bibinfo  {journal} {Curr. Opin. Solid State Mater.
  Sci.}\ }\textbf {\bibinfo {volume} {24}},\ \bibinfo {pages} {100853}
  (\bibinfo {year} {2020})}\BibitemShut {NoStop}%
\bibitem [{\citenamefont {Huang}\ \emph {et~al.}(2021)\citenamefont {Huang},
  \citenamefont {Chen},\ and\ \citenamefont {Huo}}]{HCH2021}%
  \BibitemOpen
  \bibfield  {author} {\bibinfo {author} {\bibfnamefont {H.}~\bibnamefont
  {Huang}}, \bibinfo {author} {\bibfnamefont {J.}~\bibnamefont {Chen}}, \ and\
  \bibinfo {author} {\bibfnamefont {S.}~\bibnamefont {Huo}},\ }\bibfield
  {title} {\enquote {\bibinfo {title} {Recent advances in topological elastic
  metamaterials},}\ }\href {\doibase 10.1088/1361-648X/ac27d8} {\bibfield
  {journal} {\bibinfo  {journal} {J. Condens. Matter Phys.}\ }\textbf {\bibinfo
  {volume} {33}},\ \bibinfo {pages} {503002} (\bibinfo {year}
  {2021})}\BibitemShut {NoStop}%
\bibitem [{\citenamefont {Shah}\ \emph {et~al.}(2022)\citenamefont {Shah},
  \citenamefont {Brendel}, \citenamefont {Peano},\ and\ \citenamefont
  {Marquardt}}]{SBPM2022}%
  \BibitemOpen
  \bibfield  {author} {\bibinfo {author} {\bibfnamefont {T.}~\bibnamefont
  {Shah}}, \bibinfo {author} {\bibfnamefont {C.}~\bibnamefont {Brendel}},
  \bibinfo {author} {\bibfnamefont {V.}~\bibnamefont {Peano}}, \ and\ \bibinfo
  {author} {\bibfnamefont {F.}~\bibnamefont {Marquardt}},\ }\bibfield  {title}
  {\enquote {\bibinfo {title} {Topologically protected transport in engineered
  mechanical systems},}\ }\href {https://doi.org/10.48550/arXiv.2206.1233}
  {\bibfield  {journal} {\bibinfo  {journal} {arXiv:2206.12337}\ } (\bibinfo
  {year} {2022})}\BibitemShut {NoStop}%
\bibitem [{\citenamefont {Meier}\ \emph {et~al.}(2018)\citenamefont {Meier},
  \citenamefont {Alex~An}, \citenamefont {Dauphin}, \citenamefont {Maffei},
  \citenamefont {Massignan}, \citenamefont {Hughes},\ and\ \citenamefont
  {Gadway}}]{MADMHG2018}%
  \BibitemOpen
  \bibfield  {author} {\bibinfo {author} {\bibfnamefont {E.~J.}\ \bibnamefont
  {Meier}}, \bibinfo {author} {\bibfnamefont {F.}~\bibnamefont {Alex~An}},
  \bibinfo {author} {\bibfnamefont {A.}~\bibnamefont {Dauphin}}, \bibinfo
  {author} {\bibfnamefont {M.}~\bibnamefont {Maffei}}, \bibinfo {author}
  {\bibfnamefont {P.}~\bibnamefont {Massignan}}, \bibinfo {author}
  {\bibfnamefont {T.~L.}\ \bibnamefont {Hughes}}, \ and\ \bibinfo {author}
  {\bibfnamefont {B.}~\bibnamefont {Gadway}},\ }\bibfield  {title} {\enquote
  {\bibinfo {title} {Observation of the topological {A}nderson insulator in
  disordered atomic wires},}\ }\href {\doibase 10.1126/science.aat3406}
  {\bibfield  {journal} {\bibinfo  {journal} {Science}\ }\textbf {\bibinfo
  {volume} {362}},\ \bibinfo {pages} {929} (\bibinfo {year}
  {2018})}\BibitemShut {NoStop}%
\bibitem [{\citenamefont {St{\"u}tzer}\ \emph {et~al.}(2018)\citenamefont
  {St{\"u}tzer}, \citenamefont {Plotnik}, \citenamefont {Lumer}, \citenamefont
  {Titum}, \citenamefont {Lindner}, \citenamefont {Segev}, \citenamefont
  {Rechtsman},\ and\ \citenamefont {Szameit}}]{SPLTLSRS2018}%
  \BibitemOpen
  \bibfield  {author} {\bibinfo {author} {\bibfnamefont {S.}~\bibnamefont
  {St{\"u}tzer}}, \bibinfo {author} {\bibfnamefont {Y.}~\bibnamefont
  {Plotnik}}, \bibinfo {author} {\bibfnamefont {Y.}~\bibnamefont {Lumer}},
  \bibinfo {author} {\bibfnamefont {P.}~\bibnamefont {Titum}}, \bibinfo
  {author} {\bibfnamefont {N.~H.}\ \bibnamefont {Lindner}}, \bibinfo {author}
  {\bibfnamefont {M.}~\bibnamefont {Segev}}, \bibinfo {author} {\bibfnamefont
  {M.~C.}\ \bibnamefont {Rechtsman}}, \ and\ \bibinfo {author} {\bibfnamefont
  {A.}~\bibnamefont {Szameit}},\ }\bibfield  {title} {\enquote {\bibinfo
  {title} {Photonic topological {A}nderson insulators},}\ }\href {\doibase
  10.1038/s41586-018-0418-2} {\bibfield  {journal} {\bibinfo  {journal}
  {Nature}\ }\textbf {\bibinfo {volume} {560}},\ \bibinfo {pages} {461}
  (\bibinfo {year} {2018})}\BibitemShut {NoStop}%
\bibitem [{\citenamefont {Liu}\ \emph {et~al.}(2020)\citenamefont {Liu},
  \citenamefont {Yang}, \citenamefont {Ren}, \citenamefont {Xue}, \citenamefont
  {Lin}, \citenamefont {Hu}, \citenamefont {Sun}, \citenamefont {Peng},
  \citenamefont {Zhou}, \citenamefont {Chong},\ and\ \citenamefont
  {Zhang}}]{LYRXLHSPZCZ2020}%
  \BibitemOpen
  \bibfield  {author} {\bibinfo {author} {\bibfnamefont {G.-G.}\ \bibnamefont
  {Liu}}, \bibinfo {author} {\bibfnamefont {Y.}~\bibnamefont {Yang}}, \bibinfo
  {author} {\bibfnamefont {X.}~\bibnamefont {Ren}}, \bibinfo {author}
  {\bibfnamefont {H.}~\bibnamefont {Xue}}, \bibinfo {author} {\bibfnamefont
  {X.}~\bibnamefont {Lin}}, \bibinfo {author} {\bibfnamefont {Y.-H.}\
  \bibnamefont {Hu}}, \bibinfo {author} {\bibfnamefont {H.-X.}\ \bibnamefont
  {Sun}}, \bibinfo {author} {\bibfnamefont {B.}~\bibnamefont {Peng}}, \bibinfo
  {author} {\bibfnamefont {P.}~\bibnamefont {Zhou}}, \bibinfo {author}
  {\bibfnamefont {Y.}~\bibnamefont {Chong}}, \ and\ \bibinfo {author}
  {\bibfnamefont {B.}~\bibnamefont {Zhang}},\ }\bibfield  {title} {\enquote
  {\bibinfo {title} {Topological {A}nderson insulator in disordered photonic
  crystals},}\ }\href {\doibase 10.1103/PhysRevLett.125.133603} {\bibfield
  {journal} {\bibinfo  {journal} {Phys. Rev. Lett.}\ }\textbf {\bibinfo
  {volume} {125}},\ \bibinfo {pages} {133603} (\bibinfo {year}
  {2020})}\BibitemShut {NoStop}%
\bibitem [{\citenamefont {Zangeneh-Nejad}\ and\ \citenamefont
  {Fleury}(2020)}]{ZF2020}%
  \BibitemOpen
  \bibfield  {author} {\bibinfo {author} {\bibfnamefont {F.}~\bibnamefont
  {Zangeneh-Nejad}}\ and\ \bibinfo {author} {\bibfnamefont {R.}~\bibnamefont
  {Fleury}},\ }\bibfield  {title} {\enquote {\bibinfo {title} {Disorder-induced
  signal filtering with topological metamaterials},}\ }\href {\doibase
  https://doi.org/10.1002/adma.202001034} {\bibfield  {journal} {\bibinfo
  {journal} {Adv. Mat.}\ }\textbf {\bibinfo {volume} {32}},\ \bibinfo {pages}
  {2001034} (\bibinfo {year} {2020})}\BibitemShut {NoStop}%
\bibitem [{\citenamefont {Lin}\ \emph {et~al.}(2022)\citenamefont {Lin},
  \citenamefont {Li}, \citenamefont {Xiao}, \citenamefont {Wang}, \citenamefont
  {Yi},\ and\ \citenamefont {Xue}}]{LLXWYX2022}%
  \BibitemOpen
  \bibfield  {author} {\bibinfo {author} {\bibfnamefont {Q.}~\bibnamefont
  {Lin}}, \bibinfo {author} {\bibfnamefont {T.}~\bibnamefont {Li}}, \bibinfo
  {author} {\bibfnamefont {L.}~\bibnamefont {Xiao}}, \bibinfo {author}
  {\bibfnamefont {K.}~\bibnamefont {Wang}}, \bibinfo {author} {\bibfnamefont
  {W.}~\bibnamefont {Yi}}, \ and\ \bibinfo {author} {\bibfnamefont
  {P.}~\bibnamefont {Xue}},\ }\bibfield  {title} {\enquote {\bibinfo {title}
  {Observation of non-{H}ermitian topological {A}nderson insulator in quantum
  dynamics},}\ }\href {\doibase 10.1038/s41467-022-30938-9} {\bibfield
  {journal} {\bibinfo  {journal} {Nat. Comm.}\ }\textbf {\bibinfo {volume}
  {13}},\ \bibinfo {pages} {1} (\bibinfo {year} {2022})}\BibitemShut {NoStop}%
\bibitem [{\citenamefont {Eisenberg}\ \emph {et~al.}(1998)\citenamefont
  {Eisenberg}, \citenamefont {Silberberg}, \citenamefont {Morandotti},
  \citenamefont {Boyd},\ and\ \citenamefont {Aitchison}}]{ESMBA1998}%
  \BibitemOpen
  \bibfield  {author} {\bibinfo {author} {\bibfnamefont {H.~S.}\ \bibnamefont
  {Eisenberg}}, \bibinfo {author} {\bibfnamefont {Y.}~\bibnamefont
  {Silberberg}}, \bibinfo {author} {\bibfnamefont {R.}~\bibnamefont
  {Morandotti}}, \bibinfo {author} {\bibfnamefont {A.~R.}\ \bibnamefont
  {Boyd}}, \ and\ \bibinfo {author} {\bibfnamefont {J.~S.}\ \bibnamefont
  {Aitchison}},\ }\bibfield  {title} {\enquote {\bibinfo {title} {Discrete
  spatial optical solitons in waveguide arrays},}\ }\href {\doibase
  10.1103/PhysRevLett.81.3383} {\bibfield  {journal} {\bibinfo  {journal}
  {Phys. Rev. Lett.}\ }\textbf {\bibinfo {volume} {81}},\ \bibinfo {pages}
  {3383} (\bibinfo {year} {1998})}\BibitemShut {NoStop}%
\bibitem [{\citenamefont {Leggett}(2001)}]{L2001}%
  \BibitemOpen
  \bibfield  {author} {\bibinfo {author} {\bibfnamefont {A.~J.}\ \bibnamefont
  {Leggett}},\ }\bibfield  {title} {\enquote {\bibinfo {title}
  {{B}ose-{E}instein condensation in the alkali gases: {S}ome fundamental
  concepts},}\ }\href {\doibase 10.1103/RevModPhys.73.307} {\bibfield
  {journal} {\bibinfo  {journal} {Rev. Mod. Phys.}\ }\textbf {\bibinfo {volume}
  {73}},\ \bibinfo {pages} {307} (\bibinfo {year} {2001})}\BibitemShut
  {NoStop}%
\bibitem [{\citenamefont {Trombettoni}\ and\ \citenamefont
  {Smerzi}(2001)}]{TS2001}%
  \BibitemOpen
  \bibfield  {author} {\bibinfo {author} {\bibfnamefont {A.}~\bibnamefont
  {Trombettoni}}\ and\ \bibinfo {author} {\bibfnamefont {A.}~\bibnamefont
  {Smerzi}},\ }\bibfield  {title} {\enquote {\bibinfo {title} {Discrete
  solitons and breathers with dilute {B}ose-{E}instein condensates},}\ }\href
  {\doibase 10.1103/PhysRevLett.86.2353} {\bibfield  {journal} {\bibinfo
  {journal} {Phys. Rev. Lett.}\ }\textbf {\bibinfo {volume} {86}},\ \bibinfo
  {pages} {2353} (\bibinfo {year} {2001})}\BibitemShut {NoStop}%
\bibitem [{\citenamefont {An}\ \emph {et~al.}(2021)\citenamefont {An},
  \citenamefont {Sundar}, \citenamefont {Hou}, \citenamefont {Luo},
  \citenamefont {Meier}, \citenamefont {Zhang}, \citenamefont {Hazzard},\ and\
  \citenamefont {Gadway}}]{ASHLMZHG2021}%
  \BibitemOpen
  \bibfield  {author} {\bibinfo {author} {\bibfnamefont {F.~A.}\ \bibnamefont
  {An}}, \bibinfo {author} {\bibfnamefont {B.}~\bibnamefont {Sundar}}, \bibinfo
  {author} {\bibfnamefont {J.}~\bibnamefont {Hou}}, \bibinfo {author}
  {\bibfnamefont {X.-W.}\ \bibnamefont {Luo}}, \bibinfo {author} {\bibfnamefont
  {E.~J.}\ \bibnamefont {Meier}}, \bibinfo {author} {\bibfnamefont
  {C.}~\bibnamefont {Zhang}}, \bibinfo {author} {\bibfnamefont {K.~R.~A.}\
  \bibnamefont {Hazzard}}, \ and\ \bibinfo {author} {\bibfnamefont
  {B.}~\bibnamefont {Gadway}},\ }\bibfield  {title} {\enquote {\bibinfo {title}
  {Nonlinear dynamics in a synthetic momentum-state lattice},}\ }\href
  {\doibase 10.1103/PhysRevLett.127.130401} {\bibfield  {journal} {\bibinfo
  {journal} {Phys. Rev. Lett.}\ }\textbf {\bibinfo {volume} {127}},\ \bibinfo
  {pages} {130401} (\bibinfo {year} {2021})}\BibitemShut {NoStop}%
\bibitem [{\citenamefont {Rechtsman}\ \emph {et~al.}(2013)\citenamefont
  {Rechtsman}, \citenamefont {Zeuner}, \citenamefont {Plotnik}, \citenamefont
  {Lumer}, \citenamefont {Podolsky}, \citenamefont {Dreisow}, \citenamefont
  {Nolte}, \citenamefont {Segev},\ and\ \citenamefont
  {Szameit}}]{RZPLPDNSS2013}%
  \BibitemOpen
  \bibfield  {author} {\bibinfo {author} {\bibfnamefont {M.~C.}\ \bibnamefont
  {Rechtsman}}, \bibinfo {author} {\bibfnamefont {J.~M.}\ \bibnamefont
  {Zeuner}}, \bibinfo {author} {\bibfnamefont {Y.}~\bibnamefont {Plotnik}},
  \bibinfo {author} {\bibfnamefont {Y.}~\bibnamefont {Lumer}}, \bibinfo
  {author} {\bibfnamefont {D.}~\bibnamefont {Podolsky}}, \bibinfo {author}
  {\bibfnamefont {F.}~\bibnamefont {Dreisow}}, \bibinfo {author} {\bibfnamefont
  {S.}~\bibnamefont {Nolte}}, \bibinfo {author} {\bibfnamefont
  {M.}~\bibnamefont {Segev}}, \ and\ \bibinfo {author} {\bibfnamefont
  {A.}~\bibnamefont {Szameit}},\ }\bibfield  {title} {\enquote {\bibinfo
  {title} {Photonic {F}loquet topological insulators},}\ }\href {\doibase
  10.1038/nature12066} {\bibfield  {journal} {\bibinfo  {journal} {Nature}\
  }\textbf {\bibinfo {volume} {496}},\ \bibinfo {pages} {196} (\bibinfo {year}
  {2013})}\BibitemShut {NoStop}%
\bibitem [{\citenamefont {Ablowitz}\ \emph {et~al.}(2014)\citenamefont
  {Ablowitz}, \citenamefont {Curtis},\ and\ \citenamefont {Ma}}]{ACM2014}%
  \BibitemOpen
  \bibfield  {author} {\bibinfo {author} {\bibfnamefont {M.~J.}\ \bibnamefont
  {Ablowitz}}, \bibinfo {author} {\bibfnamefont {Ch.~W.}\ \bibnamefont
  {Curtis}}, \ and\ \bibinfo {author} {\bibfnamefont {Y.-P.}\ \bibnamefont
  {Ma}},\ }\bibfield  {title} {\enquote {\bibinfo {title} {Linear and nonlinear
  traveling edge waves in optical honeycomb lattices},}\ }\href {\doibase
  10.1103/PhysRevA.90.023813} {\bibfield  {journal} {\bibinfo  {journal} {Phys.
  Rev. A}\ }\textbf {\bibinfo {volume} {90}},\ \bibinfo {pages} {023813}
  (\bibinfo {year} {2014})}\BibitemShut {NoStop}%
\bibitem [{\citenamefont {Chaunsali}\ and\ \citenamefont
  {Theocharis}(2019)}]{CT2019}%
  \BibitemOpen
  \bibfield  {author} {\bibinfo {author} {\bibfnamefont {R.}~\bibnamefont
  {Chaunsali}}\ and\ \bibinfo {author} {\bibfnamefont {G.}~\bibnamefont
  {Theocharis}},\ }\bibfield  {title} {\enquote {\bibinfo {title}
  {{Self-induced topological transition in phononic crystals by nonlinearity
  management}},}\ }\href {\doibase 10.1103/PhysRevB.100.014302} {\bibfield
  {journal} {\bibinfo  {journal} {Phys. Rev. B}\ }\textbf {\bibinfo {volume}
  {100}},\ \bibinfo {pages} {014302} (\bibinfo {year} {2019})}\BibitemShut
  {NoStop}%
\bibitem [{\citenamefont {Jezequel}\ and\ \citenamefont
  {Delplace}(2022)}]{JD2022}%
  \BibitemOpen
  \bibfield  {author} {\bibinfo {author} {\bibfnamefont {L.}~\bibnamefont
  {Jezequel}}\ and\ \bibinfo {author} {\bibfnamefont {P.}~\bibnamefont
  {Delplace}},\ }\bibfield  {title} {\enquote {\bibinfo {title} {Nonlinear edge
  modes from topological one-dimensional lattices},}\ }\href {\doibase
  10.1103/PhysRevB.105.035410} {\bibfield  {journal} {\bibinfo  {journal}
  {Phys. Rev. B}\ }\textbf {\bibinfo {volume} {105}},\ \bibinfo {pages}
  {035410} (\bibinfo {year} {2022})}\BibitemShut {NoStop}%
\bibitem [{\citenamefont {Johansson}(2023)}]{J2023}%
  \BibitemOpen
  \bibfield  {author} {\bibinfo {author} {\bibfnamefont {M.}~\bibnamefont
  {Johansson}},\ }\bibfield  {title} {\enquote {\bibinfo {title} {Topological
  edge breathers in a nonlinear {S}u-{S}chrieffer-{H}eeger lattice},}\ }\href
  {\doibase https://doi.org/10.1016/j.physleta.2022.128593} {\bibfield
  {journal} {\bibinfo  {journal} {Phys. Lett. A}\ }\textbf {\bibinfo {volume}
  {458}},\ \bibinfo {pages} {128593} (\bibinfo {year} {2023})}\BibitemShut
  {NoStop}%
\bibitem [{\citenamefont {Lumer}\ \emph {et~al.}(2013)\citenamefont {Lumer},
  \citenamefont {Plotnik}, \citenamefont {Rechtsman},\ and\ \citenamefont
  {Segev}}]{LPRS2013}%
  \BibitemOpen
  \bibfield  {author} {\bibinfo {author} {\bibfnamefont {Y.}~\bibnamefont
  {Lumer}}, \bibinfo {author} {\bibfnamefont {Y.}~\bibnamefont {Plotnik}},
  \bibinfo {author} {\bibfnamefont {M.~C.}\ \bibnamefont {Rechtsman}}, \ and\
  \bibinfo {author} {\bibfnamefont {M.}~\bibnamefont {Segev}},\ }\bibfield
  {title} {\enquote {\bibinfo {title} {Self-localized states in photonic
  topological insulators},}\ }\href {\doibase 10.1103/PhysRevLett.111.243905}
  {\bibfield  {journal} {\bibinfo  {journal} {Phys. Rev. Lett.}\ }\textbf
  {\bibinfo {volume} {111}},\ \bibinfo {pages} {243905} (\bibinfo {year}
  {2013})}\BibitemShut {NoStop}%
\bibitem [{\citenamefont {Hadad}\ \emph {et~al.}(2017)\citenamefont {Hadad},
  \citenamefont {Vitelli},\ and\ \citenamefont {Alu}}]{HVA2017}%
  \BibitemOpen
  \bibfield  {author} {\bibinfo {author} {\bibfnamefont {Y.}~\bibnamefont
  {Hadad}}, \bibinfo {author} {\bibfnamefont {V.}~\bibnamefont {Vitelli}}, \
  and\ \bibinfo {author} {\bibfnamefont {A.}~\bibnamefont {Alu}},\ }\bibfield
  {title} {\enquote {\bibinfo {title} {Solitons and propagating domain walls in
  topological resonator arrays},}\ }\href {\doibase
  10.1021/acsphotonics.7b00303} {\bibfield  {journal} {\bibinfo  {journal} {ACS
  Photonics}\ }\textbf {\bibinfo {volume} {4}},\ \bibinfo {pages} {1974}
  (\bibinfo {year} {2017})}\BibitemShut {NoStop}%
\bibitem [{\citenamefont {Asb{\'o}th}\ \emph {et~al.}(2016)\citenamefont
  {Asb{\'o}th}, \citenamefont {Oroszl{\'a}ny},\ and\ \citenamefont
  {P{\'a}lyi}}]{AOP2016}%
  \BibitemOpen
  \bibfield  {author} {\bibinfo {author} {\bibfnamefont {J.~K.}\ \bibnamefont
  {Asb{\'o}th}}, \bibinfo {author} {\bibfnamefont {L.}~\bibnamefont
  {Oroszl{\'a}ny}}, \ and\ \bibinfo {author} {\bibfnamefont {A.}~\bibnamefont
  {P{\'a}lyi}},\ }\href {\doibase https://doi.org/10.1007/978-3-319-25607-8}
  {\emph {\bibinfo {title} {A short course on topological insulators}}}\
  (\bibinfo  {publisher} {Springer Cham},\ \bibinfo {year} {2016})\BibitemShut
  {NoStop}%
\bibitem [{\citenamefont {Kitagawa}\ \emph {et~al.}(2012)\citenamefont
  {Kitagawa}, \citenamefont {Broome}, \citenamefont {Fedrizzi}, \citenamefont
  {Rudner}, \citenamefont {Berg}, \citenamefont {Kassal}, \citenamefont
  {Aspuru-Guzik}, \citenamefont {Demler},\ and\ \citenamefont
  {White}}]{KBFRBKADW2012}%
  \BibitemOpen
  \bibfield  {author} {\bibinfo {author} {\bibfnamefont {T.}~\bibnamefont
  {Kitagawa}}, \bibinfo {author} {\bibfnamefont {M.~A.}\ \bibnamefont
  {Broome}}, \bibinfo {author} {\bibfnamefont {A.}~\bibnamefont {Fedrizzi}},
  \bibinfo {author} {\bibfnamefont {M.~S.}\ \bibnamefont {Rudner}}, \bibinfo
  {author} {\bibfnamefont {E.}~\bibnamefont {Berg}}, \bibinfo {author}
  {\bibfnamefont {I.}~\bibnamefont {Kassal}}, \bibinfo {author} {\bibfnamefont
  {A.}~\bibnamefont {Aspuru-Guzik}}, \bibinfo {author} {\bibfnamefont
  {E.}~\bibnamefont {Demler}}, \ and\ \bibinfo {author} {\bibfnamefont {A.~G}\
  \bibnamefont {White}},\ }\bibfield  {title} {\enquote {\bibinfo {title}
  {Observation of topologically protected bound states in photonic quantum
  walks},}\ }\href {\doibase 10.1038/ncomms1872} {\bibfield  {journal}
  {\bibinfo  {journal} {Nat. Comm.}\ }\textbf {\bibinfo {volume} {3}},\
  \bibinfo {pages} {1} (\bibinfo {year} {2012})}\BibitemShut {NoStop}%
\bibitem [{\citenamefont {Meier}\ \emph {et~al.}(2016)\citenamefont {Meier},
  \citenamefont {An},\ and\ \citenamefont {Gadway}}]{MAG2016}%
  \BibitemOpen
  \bibfield  {author} {\bibinfo {author} {\bibfnamefont {E.~J.}\ \bibnamefont
  {Meier}}, \bibinfo {author} {\bibfnamefont {F.~A.}\ \bibnamefont {An}}, \
  and\ \bibinfo {author} {\bibfnamefont {B.}~\bibnamefont {Gadway}},\
  }\bibfield  {title} {\enquote {\bibinfo {title} {Observation of the
  topological soliton state in the {S}u--{S}chrieffer--{H}eeger model},}\
  }\href {\doibase 10.1038/ncomms13986} {\bibfield  {journal} {\bibinfo
  {journal} {Nat. Comm.}\ }\textbf {\bibinfo {volume} {7}},\ \bibinfo {pages}
  {1} (\bibinfo {year} {2016})}\BibitemShut {NoStop}%
\bibitem [{\citenamefont {St-Jean}\ \emph {et~al.}(2017)\citenamefont
  {St-Jean}, \citenamefont {Goblot}, \citenamefont {Galopin}, \citenamefont
  {Lema{\^\i}tre}, \citenamefont {Ozawa}, \citenamefont {Le~Gratiet},
  \citenamefont {Sagnes}, \citenamefont {Bloch},\ and\ \citenamefont
  {Amo}}]{SGGLOLSBA2017}%
  \BibitemOpen
  \bibfield  {author} {\bibinfo {author} {\bibfnamefont {P.}~\bibnamefont
  {St-Jean}}, \bibinfo {author} {\bibfnamefont {V.}~\bibnamefont {Goblot}},
  \bibinfo {author} {\bibfnamefont {E.}~\bibnamefont {Galopin}}, \bibinfo
  {author} {\bibfnamefont {A.}~\bibnamefont {Lema{\^\i}tre}}, \bibinfo {author}
  {\bibfnamefont {T.}~\bibnamefont {Ozawa}}, \bibinfo {author} {\bibfnamefont
  {L.}~\bibnamefont {Le~Gratiet}}, \bibinfo {author} {\bibfnamefont
  {I.}~\bibnamefont {Sagnes}}, \bibinfo {author} {\bibfnamefont
  {J.}~\bibnamefont {Bloch}}, \ and\ \bibinfo {author} {\bibfnamefont
  {A.}~\bibnamefont {Amo}},\ }\bibfield  {title} {\enquote {\bibinfo {title}
  {Lasing in topological edge states of a one-dimensional lattice},}\ }\href
  {\doibase 10.1038/s41566-017-0006-2} {\bibfield  {journal} {\bibinfo
  {journal} {Nat. Photonics}\ }\textbf {\bibinfo {volume} {11}},\ \bibinfo
  {pages} {651} (\bibinfo {year} {2017})}\BibitemShut {NoStop}%
\bibitem [{\citenamefont {Loring}\ and\ \citenamefont
  {Hastings}(2010)}]{LH2010}%
  \BibitemOpen
  \bibfield  {author} {\bibinfo {author} {\bibfnamefont {T.~A.}\ \bibnamefont
  {Loring}}\ and\ \bibinfo {author} {\bibfnamefont {M.~B.}\ \bibnamefont
  {Hastings}},\ }\bibfield  {title} {\enquote {\bibinfo {title} {Disordered
  topological insulators via c$^\star$-algebras},}\ }\href {\doibase
  10.1209/0295-5075/92/67004} {\bibfield  {journal} {\bibinfo  {journal}
  {Europhys. Lett.}\ }\textbf {\bibinfo {volume} {92}},\ \bibinfo {pages}
  {67004} (\bibinfo {year} {2010})}\BibitemShut {NoStop}%
\bibitem [{\citenamefont {Loring}(2015)}]{L2015}%
  \BibitemOpen
  \bibfield  {author} {\bibinfo {author} {\bibfnamefont {T.~A.}\ \bibnamefont
  {Loring}},\ }\bibfield  {title} {\enquote {\bibinfo {title} {K-theory and
  pseudospectra for topological insulators},}\ }\href {\doibase
  https://doi.org/10.1016/j.aop.2015.02.031} {\bibfield  {journal} {\bibinfo
  {journal} {Ann. Phys.}\ }\textbf {\bibinfo {volume} {356}},\ \bibinfo {pages}
  {383} (\bibinfo {year} {2015})}\BibitemShut {NoStop}%
\bibitem [{\citenamefont {Rieder}\ and\ \citenamefont
  {Brouwer}(2014)}]{RB2014}%
  \BibitemOpen
  \bibfield  {author} {\bibinfo {author} {\bibfnamefont {M.-Th.}\ \bibnamefont
  {Rieder}}\ and\ \bibinfo {author} {\bibfnamefont {P.~W.}\ \bibnamefont
  {Brouwer}},\ }\bibfield  {title} {\enquote {\bibinfo {title} {Density of
  states at disorder-induced phase transitions in a multichannel {M}ajorana
  wire},}\ }\href {\doibase 10.1103/PhysRevB.90.205404} {\bibfield  {journal}
  {\bibinfo  {journal} {Phys. Rev. B}\ }\textbf {\bibinfo {volume} {90}},\
  \bibinfo {pages} {205404} (\bibinfo {year} {2014})}\BibitemShut {NoStop}%
\bibitem [{\citenamefont {Shi}\ \emph {et~al.}(2021)\citenamefont {Shi},
  \citenamefont {Kiorpelidis}, \citenamefont {Chaunsali}, \citenamefont
  {Achilleos}, \citenamefont {Theocharis},\ and\ \citenamefont
  {Yang}}]{SKCATY2021}%
  \BibitemOpen
  \bibfield  {author} {\bibinfo {author} {\bibfnamefont {X.}~\bibnamefont
  {Shi}}, \bibinfo {author} {\bibfnamefont {I.}~\bibnamefont {Kiorpelidis}},
  \bibinfo {author} {\bibfnamefont {R.}~\bibnamefont {Chaunsali}}, \bibinfo
  {author} {\bibfnamefont {V.}~\bibnamefont {Achilleos}}, \bibinfo {author}
  {\bibfnamefont {G.}~\bibnamefont {Theocharis}}, \ and\ \bibinfo {author}
  {\bibfnamefont {J.}~\bibnamefont {Yang}},\ }\bibfield  {title} {\enquote
  {\bibinfo {title} {Disorder-induced topological phase transition in a
  one-dimensional mechanical system},}\ }\href {\doibase
  10.1103/PhysRevResearch.3.033012} {\bibfield  {journal} {\bibinfo  {journal}
  {Phys. Rev. Research}\ }\textbf {\bibinfo {volume} {3}},\ \bibinfo {pages}
  {033012} (\bibinfo {year} {2021})}\BibitemShut {NoStop}%
\bibitem [{\citenamefont {Skipetrov}\ and\ \citenamefont
  {Wulles}(2022)}]{SW2022}%
  \BibitemOpen
  \bibfield  {author} {\bibinfo {author} {\bibfnamefont {S.~E.}\ \bibnamefont
  {Skipetrov}}\ and\ \bibinfo {author} {\bibfnamefont {P.}~\bibnamefont
  {Wulles}},\ }\bibfield  {title} {\enquote {\bibinfo {title} {Topological
  transitions and {A}nderson localization of light in disordered atomic
  arrays},}\ }\href {\doibase 10.1103/PhysRevA.105.043514} {\bibfield
  {journal} {\bibinfo  {journal} {Phys. Rev. A}\ }\textbf {\bibinfo {volume}
  {105}},\ \bibinfo {pages} {043514} (\bibinfo {year} {2022})}\BibitemShut
  {NoStop}%
\bibitem [{\citenamefont {Cardano}\ \emph {et~al.}(2017)\citenamefont
  {Cardano}, \citenamefont {D’Errico}, \citenamefont {Dauphin}, \citenamefont
  {Maffei}, \citenamefont {Piccirillo}, \citenamefont {De~Lisio}, \citenamefont
  {De~Filippis}, \citenamefont {Cataudella}, \citenamefont {Santamato},
  \citenamefont {Marrucci}, \citenamefont {Lewenstein},\ and\ \citenamefont
  {Massignan}}]{CDDMPDDCSMLM2017}%
  \BibitemOpen
  \bibfield  {author} {\bibinfo {author} {\bibfnamefont {F.}~\bibnamefont
  {Cardano}}, \bibinfo {author} {\bibfnamefont {A.}~\bibnamefont {D’Errico}},
  \bibinfo {author} {\bibfnamefont {A.}~\bibnamefont {Dauphin}}, \bibinfo
  {author} {\bibfnamefont {M.}~\bibnamefont {Maffei}}, \bibinfo {author}
  {\bibfnamefont {B.}~\bibnamefont {Piccirillo}}, \bibinfo {author}
  {\bibfnamefont {C.}~\bibnamefont {De~Lisio}}, \bibinfo {author}
  {\bibfnamefont {G.}~\bibnamefont {De~Filippis}}, \bibinfo {author}
  {\bibfnamefont {V.}~\bibnamefont {Cataudella}}, \bibinfo {author}
  {\bibfnamefont {E.}~\bibnamefont {Santamato}}, \bibinfo {author}
  {\bibfnamefont {L.}~\bibnamefont {Marrucci}}, \bibinfo {author}
  {\bibfnamefont {M.}~\bibnamefont {Lewenstein}}, \ and\ \bibinfo {author}
  {\bibfnamefont {P.}~\bibnamefont {Massignan}},\ }\bibfield  {title} {\enquote
  {\bibinfo {title} {Detection of {Z}ak phases and topological invariants in a
  chiral quantum walk of twisted photons},}\ }\href {\doibase
  10.1038/ncomms15516} {\bibfield  {journal} {\bibinfo  {journal} {Nat. Comm.}\
  }\textbf {\bibinfo {volume} {8}},\ \bibinfo {pages} {1} (\bibinfo {year}
  {2017})}\BibitemShut {NoStop}%
\bibitem [{\citenamefont {Maffei}(2019)}]{M2019}%
  \BibitemOpen
  \bibfield  {author} {\bibinfo {author} {\bibfnamefont {M.}~\bibnamefont
  {Maffei}},\ }\emph {\bibinfo {title} {Simulation and bulk detection of
  topological phases of matter}},\ \href {http://hdl.handle.net/2117/129268}
  {Ph.D. thesis},\ \bibinfo {address} {Polytechnic University of Catalonia}
  (\bibinfo {year} {2019})\BibitemShut {NoStop}%
\bibitem [{\citenamefont {Wang}\ \emph {et~al.}(2018)\citenamefont {Wang},
  \citenamefont {Xiao}, \citenamefont {Qiu}, \citenamefont {Wang},
  \citenamefont {Yi},\ and\ \citenamefont {Xue}}]{WXQWYX2018}%
  \BibitemOpen
  \bibfield  {author} {\bibinfo {author} {\bibfnamefont {X.}~\bibnamefont
  {Wang}}, \bibinfo {author} {\bibfnamefont {L.}~\bibnamefont {Xiao}}, \bibinfo
  {author} {\bibfnamefont {X.}~\bibnamefont {Qiu}}, \bibinfo {author}
  {\bibfnamefont {K.}~\bibnamefont {Wang}}, \bibinfo {author} {\bibfnamefont
  {W.}~\bibnamefont {Yi}}, \ and\ \bibinfo {author} {\bibfnamefont
  {P.}~\bibnamefont {Xue}},\ }\bibfield  {title} {\enquote {\bibinfo {title}
  {Detecting topological invariants and revealing topological phase transitions
  in discrete-time photonic quantum walks},}\ }\href {\doibase
  10.1103/PhysRevA.98.013835} {\bibfield  {journal} {\bibinfo  {journal} {Phys.
  Rev. A}\ }\textbf {\bibinfo {volume} {98}},\ \bibinfo {pages} {013835}
  (\bibinfo {year} {2018})}\BibitemShut {NoStop}%
\bibitem [{\citenamefont {Wang}\ \emph {et~al.}(2019)\citenamefont {Wang},
  \citenamefont {Lu}, \citenamefont {Mei}, \citenamefont {Gao}, \citenamefont
  {Li}, \citenamefont {Tang}, \citenamefont {Zhu}, \citenamefont {Jia},\ and\
  \citenamefont {Jin}}]{WLMGLTZJJ2019}%
  \BibitemOpen
  \bibfield  {author} {\bibinfo {author} {\bibfnamefont {Y.}~\bibnamefont
  {Wang}}, \bibinfo {author} {\bibfnamefont {Y.-H.}\ \bibnamefont {Lu}},
  \bibinfo {author} {\bibfnamefont {F.}~\bibnamefont {Mei}}, \bibinfo {author}
  {\bibfnamefont {J.}~\bibnamefont {Gao}}, \bibinfo {author} {\bibfnamefont
  {Z.-M.}\ \bibnamefont {Li}}, \bibinfo {author} {\bibfnamefont
  {H.}~\bibnamefont {Tang}}, \bibinfo {author} {\bibfnamefont {S.-L.}\
  \bibnamefont {Zhu}}, \bibinfo {author} {\bibfnamefont {S.}~\bibnamefont
  {Jia}}, \ and\ \bibinfo {author} {\bibfnamefont {X.-M.}\ \bibnamefont
  {Jin}},\ }\bibfield  {title} {\enquote {\bibinfo {title} {Direct observation
  of topology from single-photon dynamics},}\ }\href {\doibase
  10.1103/PhysRevLett.122.193903} {\bibfield  {journal} {\bibinfo  {journal}
  {Phys. Rev. Lett.}\ }\textbf {\bibinfo {volume} {122}},\ \bibinfo {pages}
  {193903} (\bibinfo {year} {2019})}\BibitemShut {NoStop}%
\bibitem [{\citenamefont {Xie}\ \emph {et~al.}(2020)\citenamefont {Xie},
  \citenamefont {Deng}, \citenamefont {Xiao}, \citenamefont {Gou},
  \citenamefont {Chen}, \citenamefont {Yi},\ and\ \citenamefont
  {Yan}}]{XDXGCYY2020}%
  \BibitemOpen
  \bibfield  {author} {\bibinfo {author} {\bibfnamefont {D.}~\bibnamefont
  {Xie}}, \bibinfo {author} {\bibfnamefont {T.-S.}\ \bibnamefont {Deng}},
  \bibinfo {author} {\bibfnamefont {T.}~\bibnamefont {Xiao}}, \bibinfo {author}
  {\bibfnamefont {W.}~\bibnamefont {Gou}}, \bibinfo {author} {\bibfnamefont
  {T.}~\bibnamefont {Chen}}, \bibinfo {author} {\bibfnamefont {W.}~\bibnamefont
  {Yi}}, \ and\ \bibinfo {author} {\bibfnamefont {B.}~\bibnamefont {Yan}},\
  }\bibfield  {title} {\enquote {\bibinfo {title} {Topological quantum walks in
  momentum space with a {B}ose-{E}instein condensate},}\ }\href {\doibase
  10.1103/PhysRevLett.124.050502} {\bibfield  {journal} {\bibinfo  {journal}
  {Phys. Rev. Lett.}\ }\textbf {\bibinfo {volume} {124}},\ \bibinfo {pages}
  {050502} (\bibinfo {year} {2020})}\BibitemShut {NoStop}%
\bibitem [{\citenamefont {Su}\ \emph {et~al.}(1979)\citenamefont {Su},
  \citenamefont {Schrieffer},\ and\ \citenamefont {Heeger}}]{SSH1979}%
  \BibitemOpen
  \bibfield  {author} {\bibinfo {author} {\bibfnamefont {W.~P.}\ \bibnamefont
  {Su}}, \bibinfo {author} {\bibfnamefont {J.~R.}\ \bibnamefont {Schrieffer}},
  \ and\ \bibinfo {author} {\bibfnamefont {A.~J.}\ \bibnamefont {Heeger}},\
  }\bibfield  {title} {\enquote {\bibinfo {title} {Solitons in
  polyacetylene},}\ }\href {\doibase 10.1103/PhysRevLett.42.1698} {\bibfield
  {journal} {\bibinfo  {journal} {Phys. Rev. Lett.}\ }\textbf {\bibinfo
  {volume} {42}},\ \bibinfo {pages} {1698} (\bibinfo {year}
  {1979})}\BibitemShut {NoStop}%
\bibitem [{\citenamefont {Batra}\ and\ \citenamefont {Sheet}(2020)}]{BS2020}%
  \BibitemOpen
  \bibfield  {author} {\bibinfo {author} {\bibfnamefont {N.}~\bibnamefont
  {Batra}}\ and\ \bibinfo {author} {\bibfnamefont {G.}~\bibnamefont {Sheet}},\
  }\bibfield  {title} {\enquote {\bibinfo {title} {Physics with coffee and
  doughnuts},}\ }\href {\doibase 10.1007/s12045-020-0995-x} {\bibfield
  {journal} {\bibinfo  {journal} {Reson.}\ }\textbf {\bibinfo {volume} {25}},\
  \bibinfo {pages} {765} (\bibinfo {year} {2020})}\BibitemShut {NoStop}%
\bibitem [{\citenamefont {Krapivsky}\ and\ \citenamefont
  {Luck}(2011)}]{KL2011}%
  \BibitemOpen
  \bibfield  {author} {\bibinfo {author} {\bibfnamefont {P.~L.}\ \bibnamefont
  {Krapivsky}}\ and\ \bibinfo {author} {\bibfnamefont {J.~M.}\ \bibnamefont
  {Luck}},\ }\bibfield  {title} {\enquote {\bibinfo {title} {Dynamics of a
  quantum particle in low-dimensional disordered systems with extended
  states},}\ }\href {\doibase 10.1088/1742-5468/2011/02/p02031} {\bibfield
  {journal} {\bibinfo  {journal} {J. Stat. Mech. Theory Exp.}\ }\textbf
  {\bibinfo {volume} {2011}},\ \bibinfo {pages} {P02031} (\bibinfo {year}
  {2011})}\BibitemShut {NoStop}%
\bibitem [{\citenamefont {Mondragon-Shem}\ \emph {et~al.}(2014)\citenamefont
  {Mondragon-Shem}, \citenamefont {Hughes}, \citenamefont {Song},\ and\
  \citenamefont {Prodan}}]{MHSP2014}%
  \BibitemOpen
  \bibfield  {author} {\bibinfo {author} {\bibfnamefont {I.}~\bibnamefont
  {Mondragon-Shem}}, \bibinfo {author} {\bibfnamefont {T.~L.}\ \bibnamefont
  {Hughes}}, \bibinfo {author} {\bibfnamefont {J.}~\bibnamefont {Song}}, \ and\
  \bibinfo {author} {\bibfnamefont {E.}~\bibnamefont {Prodan}},\ }\bibfield
  {title} {\enquote {\bibinfo {title} {Topological criticality in the
  chiral-symmetric {AIII} class at strong disorder},}\ }\href {\doibase
  10.1103/PhysRevLett.113.046802} {\bibfield  {journal} {\bibinfo  {journal}
  {Phys. Rev. Lett.}\ }\textbf {\bibinfo {volume} {113}},\ \bibinfo {pages}
  {046802} (\bibinfo {year} {2014})}\BibitemShut {NoStop}%
\bibitem [{\citenamefont {Inui}\ \emph {et~al.}(1994)\citenamefont {Inui},
  \citenamefont {Trugman},\ and\ \citenamefont {Abrahams}}]{ITA1994}%
  \BibitemOpen
  \bibfield  {author} {\bibinfo {author} {\bibfnamefont {M.}~\bibnamefont
  {Inui}}, \bibinfo {author} {\bibfnamefont {S.~A.}\ \bibnamefont {Trugman}}, \
  and\ \bibinfo {author} {\bibfnamefont {E.}~\bibnamefont {Abrahams}},\
  }\bibfield  {title} {\enquote {\bibinfo {title} {Unusual properties of
  midband states in systems with off-diagonal disorder},}\ }\href {\doibase
  10.1103/PhysRevB.49.3190} {\bibfield  {journal} {\bibinfo  {journal} {Phys.
  Rev. B}\ }\textbf {\bibinfo {volume} {49}},\ \bibinfo {pages} {3190}
  (\bibinfo {year} {1994})}\BibitemShut {NoStop}%
\bibitem [{\citenamefont {Izrailev}\ \emph {et~al.}(2012)\citenamefont
  {Izrailev}, \citenamefont {Krokhin},\ and\ \citenamefont
  {Makarov}}]{IKM2012}%
  \BibitemOpen
  \bibfield  {author} {\bibinfo {author} {\bibfnamefont {F.~M.}\ \bibnamefont
  {Izrailev}}, \bibinfo {author} {\bibfnamefont {A.~A.}\ \bibnamefont
  {Krokhin}}, \ and\ \bibinfo {author} {\bibfnamefont {N.~M.}\ \bibnamefont
  {Makarov}},\ }\bibfield  {title} {\enquote {\bibinfo {title} {Anomalous
  localization in low-dimensional systems with correlated disorder},}\ }\href
  {\doibase https://doi.org/10.1016/j.physrep.2011.11.002} {\bibfield
  {journal} {\bibinfo  {journal} {Phys. Rep.}\ }\textbf {\bibinfo {volume}
  {512}},\ \bibinfo {pages} {125} (\bibinfo {year} {2012})}\BibitemShut
  {NoStop}%
\bibitem [{\citenamefont {Bianco}\ and\ \citenamefont {Resta}(2011)}]{BR2011}%
  \BibitemOpen
  \bibfield  {author} {\bibinfo {author} {\bibfnamefont {R.}~\bibnamefont
  {Bianco}}\ and\ \bibinfo {author} {\bibfnamefont {R.}~\bibnamefont {Resta}},\
  }\bibfield  {title} {\enquote {\bibinfo {title} {Mapping topological order in
  coordinate space},}\ }\href {\doibase 10.1103/PhysRevB.84.241106} {\bibfield
  {journal} {\bibinfo  {journal} {Phys. Rev. B}\ }\textbf {\bibinfo {volume}
  {84}},\ \bibinfo {pages} {241106} (\bibinfo {year} {2011})}\BibitemShut
  {NoStop}%
\bibitem [{\citenamefont {Bianco}\ and\ \citenamefont {Resta}(2013)}]{BR2013}%
  \BibitemOpen
  \bibfield  {author} {\bibinfo {author} {\bibfnamefont {R.}~\bibnamefont
  {Bianco}}\ and\ \bibinfo {author} {\bibfnamefont {R.}~\bibnamefont {Resta}},\
  }\bibfield  {title} {\enquote {\bibinfo {title} {Orbital magnetization as a
  local property},}\ }\href {\doibase 10.1103/PhysRevLett.110.087202}
  {\bibfield  {journal} {\bibinfo  {journal} {Phys. Rev. Lett.}\ }\textbf
  {\bibinfo {volume} {110}},\ \bibinfo {pages} {087202} (\bibinfo {year}
  {2013})}\BibitemShut {NoStop}%
\bibitem [{\citenamefont {Fleishman}\ and\ \citenamefont
  {Licciardello}(1977)}]{FL1977}%
  \BibitemOpen
  \bibfield  {author} {\bibinfo {author} {\bibfnamefont {L.}~\bibnamefont
  {Fleishman}}\ and\ \bibinfo {author} {\bibfnamefont {D.~C.}\ \bibnamefont
  {Licciardello}},\ }\bibfield  {title} {\enquote {\bibinfo {title}
  {Fluctuations and localization in one dimension},}\ }\href {\doibase
  10.1088/0022-3719/10/6/003} {\bibfield  {journal} {\bibinfo  {journal} {J.
  Phys. C: Solid State Phys.}\ }\textbf {\bibinfo {volume} {10}},\ \bibinfo
  {pages} {L125} (\bibinfo {year} {1977})}\BibitemShut {NoStop}%
\bibitem [{\citenamefont {Soukoulis}\ and\ \citenamefont
  {Economou}(1981)}]{SE1981}%
  \BibitemOpen
  \bibfield  {author} {\bibinfo {author} {\bibfnamefont {C.~M.}\ \bibnamefont
  {Soukoulis}}\ and\ \bibinfo {author} {\bibfnamefont {E.~N.}\ \bibnamefont
  {Economou}},\ }\bibfield  {title} {\enquote {\bibinfo {title} {Off-diagonal
  disorder in one-dimensional systems},}\ }\href {\doibase
  10.1103/PhysRevB.24.5698} {\bibfield  {journal} {\bibinfo  {journal} {Phys.
  Rev. B}\ }\textbf {\bibinfo {volume} {24}},\ \bibinfo {pages} {5698}
  (\bibinfo {year} {1981})}\BibitemShut {NoStop}%
\bibitem [{\citenamefont {Dyson}(1953)}]{D1953}%
  \BibitemOpen
  \bibfield  {author} {\bibinfo {author} {\bibfnamefont {F.~J.}\ \bibnamefont
  {Dyson}},\ }\bibfield  {title} {\enquote {\bibinfo {title} {The dynamics of a
  disordered linear chain},}\ }\href {\doibase 10.1103/PhysRev.92.1331}
  {\bibfield  {journal} {\bibinfo  {journal} {Phys. Rev.}\ }\textbf {\bibinfo
  {volume} {92}},\ \bibinfo {pages} {1331} (\bibinfo {year}
  {1953})}\BibitemShut {NoStop}%
\bibitem [{NOT()}]{NOTE2022a}%
  \BibitemOpen
  \href@noop {} {}\bibinfo {note} {The eigenenergies are obtained by numerical
  diagonalization of the Hamiltonian matrix $\mathbb{H}$
  [Eq.~\eqref{eq:eigenvalue}] with periodic boundary conditions at the edges of
  the chain in order to mitigate finite size effects. The same computations
  performed using open boundary conditions, show similar tendencies, but at
  larger lattice sizes~\cite{KL2011}.}\BibitemShut {Stop}%
\bibitem [{\citenamefont {Hairer}\ \emph {et~al.}(2006)\citenamefont {Hairer},
  \citenamefont {Wanner},\ and\ \citenamefont {Lubich}}]{HWL2006}%
  \BibitemOpen
  \bibfield  {author} {\bibinfo {author} {\bibfnamefont {E.}~\bibnamefont
  {Hairer}}, \bibinfo {author} {\bibfnamefont {G.}~\bibnamefont {Wanner}}, \
  and\ \bibinfo {author} {\bibfnamefont {C.}~\bibnamefont {Lubich}},\ }\href
  {\doibase https://doi.org/10.1007/3-540-30666-8} {\emph {\bibinfo {title}
  {Geometric Numerical Integration: Structure-Preserving Algorithms for
  Ordinary Differential Equations}}}\ (\bibinfo  {publisher} {Springer Berlin
  Heidelberg},\ \bibinfo {year} {2006})\BibitemShut {NoStop}%
\bibitem [{\citenamefont {Blanes}\ \emph {et~al.}(2013)\citenamefont {Blanes},
  \citenamefont {Casas}, \citenamefont {Farres}, \citenamefont {Laskar},
  \citenamefont {Makazaga},\ and\ \citenamefont {Murua}}]{BCFLMM2013}%
  \BibitemOpen
  \bibfield  {author} {\bibinfo {author} {\bibfnamefont {S.}~\bibnamefont
  {Blanes}}, \bibinfo {author} {\bibfnamefont {F.}~\bibnamefont {Casas}},
  \bibinfo {author} {\bibfnamefont {A.}~\bibnamefont {Farres}}, \bibinfo
  {author} {\bibfnamefont {J.}~\bibnamefont {Laskar}}, \bibinfo {author}
  {\bibfnamefont {J.}~\bibnamefont {Makazaga}}, \ and\ \bibinfo {author}
  {\bibfnamefont {A.}~\bibnamefont {Murua}},\ }\bibfield  {title} {\enquote
  {\bibinfo {title} {New families of symplectic splitting methods for numerical
  integration in dynamical astronomy},}\ }\href {\doibase
  10.1016/j.apnum.2013.01.003} {\bibfield  {journal} {\bibinfo  {journal}
  {Appl. Num. Math.}\ }\textbf {\bibinfo {volume} {68}},\ \bibinfo {pages} {58}
  (\bibinfo {year} {2013})}\BibitemShut {NoStop}%
\bibitem [{\citenamefont {Skokos}\ \emph {et~al.}(2014)\citenamefont {Skokos},
  \citenamefont {Gerlach}, \citenamefont {Bodyfelt}, \citenamefont
  {Papamikos},\ and\ \citenamefont {Eggl}}]{SGBPE2014}%
  \BibitemOpen
  \bibfield  {author} {\bibinfo {author} {\bibfnamefont {Ch.}\ \bibnamefont
  {Skokos}}, \bibinfo {author} {\bibfnamefont {E.}~\bibnamefont {Gerlach}},
  \bibinfo {author} {\bibfnamefont {J.D.}\ \bibnamefont {Bodyfelt}}, \bibinfo
  {author} {\bibfnamefont {G.}~\bibnamefont {Papamikos}}, \ and\ \bibinfo
  {author} {\bibfnamefont {S.}~\bibnamefont {Eggl}},\ }\bibfield  {title}
  {\enquote {\bibinfo {title} {High order three part split symplectic
  integrators: {E}fficient techniques for the long time simulation of the
  disordered discrete nonlinear {S}chr\"odinger equation},}\ }\href {\doibase
  https://doi.org/10.1016/j.physleta.2014.04.050} {\bibfield  {journal}
  {\bibinfo  {journal} {Phys. Lett. A}\ }\textbf {\bibinfo {volume} {378}},\
  \bibinfo {pages} {1809} (\bibinfo {year} {2014})}\BibitemShut {NoStop}%
\bibitem [{\citenamefont {Gerlach}\ \emph {et~al.}(2016)\citenamefont
  {Gerlach}, \citenamefont {Meichsner},\ and\ \citenamefont
  {Skokos}}]{GMS2016}%
  \BibitemOpen
  \bibfield  {author} {\bibinfo {author} {\bibfnamefont {E.}~\bibnamefont
  {Gerlach}}, \bibinfo {author} {\bibfnamefont {J.}~\bibnamefont {Meichsner}},
  \ and\ \bibinfo {author} {\bibfnamefont {Ch.}\ \bibnamefont {Skokos}},\
  }\bibfield  {title} {\enquote {\bibinfo {title} {On the symplectic
  integration of the discrete nonlinear {S}chr{\"o}dinger equation with
  disorder},}\ }\href {\doibase 10.1140/epjst/e2016-02657-0} {\bibfield
  {journal} {\bibinfo  {journal} {Eur. Phys. J. Spec. Top.}\ }\textbf {\bibinfo
  {volume} {225}},\ \bibinfo {pages} {1103} (\bibinfo {year}
  {2016})}\BibitemShut {NoStop}%
\bibitem [{\citenamefont {Danieli}\ \emph {et~al.}(2019)\citenamefont
  {Danieli}, \citenamefont {Many~Manda}, \citenamefont {Mithun},\ and\
  \citenamefont {Skokos}}]{DMMS2019}%
  \BibitemOpen
  \bibfield  {author} {\bibinfo {author} {\bibfnamefont {C.}~\bibnamefont
  {Danieli}}, \bibinfo {author} {\bibfnamefont {B.}~\bibnamefont {Many~Manda}},
  \bibinfo {author} {\bibfnamefont {T.}~\bibnamefont {Mithun}}, \ and\ \bibinfo
  {author} {\bibfnamefont {Ch.}\ \bibnamefont {Skokos}},\ }\bibfield  {title}
  {\enquote {\bibinfo {title} {Computational efficiency of numerical
  integration methods for the tangent dynamics of many-body {H}amiltonian
  systems in one and two spatial dimensions},}\ }\href {\doibase
  10.3934/mine.2019.3.447} {\bibfield  {journal} {\bibinfo  {journal} {Math.
  Eng.}\ }\textbf {\bibinfo {volume} {1}},\ \bibinfo {pages} {447} (\bibinfo
  {year} {2019})}\BibitemShut {NoStop}%
\bibitem [{\citenamefont {Lepri}\ \emph {et~al.}(2010)\citenamefont {Lepri},
  \citenamefont {Schilling},\ and\ \citenamefont {Aubry}}]{LSA2010}%
  \BibitemOpen
  \bibfield  {author} {\bibinfo {author} {\bibfnamefont {S.}~\bibnamefont
  {Lepri}}, \bibinfo {author} {\bibfnamefont {R.}~\bibnamefont {Schilling}}, \
  and\ \bibinfo {author} {\bibfnamefont {S.}~\bibnamefont {Aubry}},\ }\bibfield
   {title} {\enquote {\bibinfo {title} {Asymptotic energy profile of a wave
  packet in disordered chains},}\ }\href {\doibase 10.1103/PhysRevE.82.056602}
  {\bibfield  {journal} {\bibinfo  {journal} {Phys. Rev. E}\ }\textbf {\bibinfo
  {volume} {82}},\ \bibinfo {pages} {056602} (\bibinfo {year}
  {2010})}\BibitemShut {NoStop}%
\bibitem [{\citenamefont {Achilleos}\ \emph {et~al.}(2016)\citenamefont
  {Achilleos}, \citenamefont {Theocharis},\ and\ \citenamefont
  {Skokos}}]{ATS2016}%
  \BibitemOpen
  \bibfield  {author} {\bibinfo {author} {\bibfnamefont {V.}~\bibnamefont
  {Achilleos}}, \bibinfo {author} {\bibfnamefont {G.}~\bibnamefont
  {Theocharis}}, \ and\ \bibinfo {author} {\bibfnamefont {Ch.}\ \bibnamefont
  {Skokos}},\ }\bibfield  {title} {\enquote {\bibinfo {title} {Energy transport
  in one-dimensional disordered granular solids},}\ }\href {\doibase
  10.1103/PhysRevE.93.022903} {\bibfield  {journal} {\bibinfo  {journal} {Phys.
  Rev. E}\ }\textbf {\bibinfo {volume} {93}},\ \bibinfo {pages} {022903}
  (\bibinfo {year} {2016})}\BibitemShut {NoStop}%
\bibitem [{\citenamefont {Anderson}(1958)}]{A1958}%
  \BibitemOpen
  \bibfield  {author} {\bibinfo {author} {\bibfnamefont {P.~W.}\ \bibnamefont
  {Anderson}},\ }\bibfield  {title} {\enquote {\bibinfo {title} {Absence of
  diffusion in certain random lattices},}\ }\href {\doibase
  10.1103/PhysRev.109.1492} {\bibfield  {journal} {\bibinfo  {journal} {Phys.
  Rev.}\ }\textbf {\bibinfo {volume} {109}},\ \bibinfo {pages} {1492} (\bibinfo
  {year} {1958})}\BibitemShut {NoStop}%
\bibitem [{\citenamefont {Kramer}\ and\ \citenamefont
  {MacKinnon}(1993)}]{KM1993}%
  \BibitemOpen
  \bibfield  {author} {\bibinfo {author} {\bibfnamefont {B.}~\bibnamefont
  {Kramer}}\ and\ \bibinfo {author} {\bibfnamefont {A.}~\bibnamefont
  {MacKinnon}},\ }\bibfield  {title} {\enquote {\bibinfo {title} {Localization:
  {T}heory and experiment},}\ }\href {\doibase 10.1088/0034-4885/56/12/001}
  {\bibfield  {journal} {\bibinfo  {journal} {Rep. Prog. Phys.}\ }\textbf
  {\bibinfo {volume} {56}},\ \bibinfo {pages} {1469} (\bibinfo {year}
  {1993})}\BibitemShut {NoStop}%
\bibitem [{\citenamefont {Krishna}\ and\ \citenamefont {Bhatt}(2021)}]{KB2021}%
  \BibitemOpen
  \bibfield  {author} {\bibinfo {author} {\bibfnamefont {A.}~\bibnamefont
  {Krishna}}\ and\ \bibinfo {author} {\bibfnamefont {R.N.}\ \bibnamefont
  {Bhatt}},\ }\bibfield  {title} {\enquote {\bibinfo {title} {Beyond the
  universal {D}yson singularity for {1-D} chains with hopping disorder},}\
  }\href {\doibase https://doi.org/10.1016/j.aop.2021.168537} {\bibfield
  {journal} {\bibinfo  {journal} {Ann. Phys.}\ }\textbf {\bibinfo {volume}
  {435}},\ \bibinfo {pages} {168537} (\bibinfo {year} {2021})}\BibitemShut
  {NoStop}%
\bibitem [{\citenamefont {Ma}\ and\ \citenamefont {Susanto}(2021)}]{MS2021}%
  \BibitemOpen
  \bibfield  {author} {\bibinfo {author} {\bibfnamefont {Y.-P.}\ \bibnamefont
  {Ma}}\ and\ \bibinfo {author} {\bibfnamefont {H.}~\bibnamefont {Susanto}},\
  }\bibfield  {title} {\enquote {\bibinfo {title} {Topological edge solitons
  and their stability in a nonlinear {S}u-{S}chrieffer-{H}eeger model},}\
  }\href {\doibase 10.1103/PhysRevE.104.054206} {\bibfield  {journal} {\bibinfo
   {journal} {Phys. Rev. E}\ }\textbf {\bibinfo {volume} {104}},\ \bibinfo
  {pages} {054206} (\bibinfo {year} {2021})}\BibitemShut {NoStop}%
\bibitem [{\citenamefont {Skokos}\ \emph {et~al.}(2009)\citenamefont {Skokos},
  \citenamefont {Krimer}, \citenamefont {Komineas},\ and\ \citenamefont
  {Flach}}]{SKKF2009}%
  \BibitemOpen
  \bibfield  {author} {\bibinfo {author} {\bibfnamefont {Ch.}\ \bibnamefont
  {Skokos}}, \bibinfo {author} {\bibfnamefont {D.~O.}\ \bibnamefont {Krimer}},
  \bibinfo {author} {\bibfnamefont {S.}~\bibnamefont {Komineas}}, \ and\
  \bibinfo {author} {\bibfnamefont {S.}~\bibnamefont {Flach}},\ }\bibfield
  {title} {\enquote {\bibinfo {title} {Delocalization of wave packets in
  disordered nonlinear chains},}\ }\href {\doibase 10.1103/PhysRevE.79.056211}
  {\bibfield  {journal} {\bibinfo  {journal} {Phys. Rev. E}\ }\textbf {\bibinfo
  {volume} {79}},\ \bibinfo {pages} {056211} (\bibinfo {year}
  {2009})}\BibitemShut {NoStop}%
\bibitem [{\citenamefont {Ermann}\ and\ \citenamefont
  {Shepelyansky}(2021)}]{ES2021}%
  \BibitemOpen
  \bibfield  {author} {\bibinfo {author} {\bibfnamefont {L.}~\bibnamefont
  {Ermann}}\ and\ \bibinfo {author} {\bibfnamefont {D.~L.}\ \bibnamefont
  {Shepelyansky}},\ }\bibfield  {title} {\enquote {\bibinfo {title}
  {Deconfinement of classical {Y}ang–{M}ills color fields in a disorder
  potential},}\ }\href {\doibase 10.1063/5.0057969} {\bibfield  {journal}
  {\bibinfo  {journal} {Chaos}\ }\textbf {\bibinfo {volume} {31}},\ \bibinfo
  {pages} {093106} (\bibinfo {year} {2021})}\BibitemShut {NoStop}%
\bibitem [{\citenamefont {Shepelyansky}(1993)}]{S1993}%
  \BibitemOpen
  \bibfield  {author} {\bibinfo {author} {\bibfnamefont {D.~L.}\ \bibnamefont
  {Shepelyansky}},\ }\bibfield  {title} {\enquote {\bibinfo {title}
  {Delocalization of quantum chaos by weak nonlinearity},}\ }\href {\doibase
  10.1103/PhysRevLett.70.1787} {\bibfield  {journal} {\bibinfo  {journal}
  {Phys. Rev. Lett.}\ }\textbf {\bibinfo {volume} {70}},\ \bibinfo {pages}
  {1787} (\bibinfo {year} {1993})}\BibitemShut {NoStop}%
\bibitem [{\citenamefont {Flach}\ \emph {et~al.}(2009)\citenamefont {Flach},
  \citenamefont {Krimer},\ and\ \citenamefont {Skokos}}]{FKS2009}%
  \BibitemOpen
  \bibfield  {author} {\bibinfo {author} {\bibfnamefont {S.}~\bibnamefont
  {Flach}}, \bibinfo {author} {\bibfnamefont {D.~O.}\ \bibnamefont {Krimer}}, \
  and\ \bibinfo {author} {\bibfnamefont {Ch.}\ \bibnamefont {Skokos}},\
  }\bibfield  {title} {\enquote {\bibinfo {title} {Universal spreading of wave
  packets in disordered nonlinear systems},}\ }\href
  {https://link.aps.org/doi/10.1103/PhysRevLett.102.024101} {\bibfield
  {journal} {\bibinfo  {journal} {Phys. Rev. Lett.}\ }\textbf {\bibinfo
  {volume} {102}},\ \bibinfo {pages} {024101} (\bibinfo {year}
  {2009})}\BibitemShut {NoStop}%
\bibitem [{\citenamefont {Skokos}\ \emph {et~al.}(2013)\citenamefont {Skokos},
  \citenamefont {Gkolias},\ and\ \citenamefont {Flach}}]{SGF2013}%
  \BibitemOpen
  \bibfield  {author} {\bibinfo {author} {\bibfnamefont {Ch.}\ \bibnamefont
  {Skokos}}, \bibinfo {author} {\bibfnamefont {I.}~\bibnamefont {Gkolias}}, \
  and\ \bibinfo {author} {\bibfnamefont {S.}~\bibnamefont {Flach}},\ }\bibfield
   {title} {\enquote {\bibinfo {title} {Nonequilibrium chaos of disordered
  nonlinear waves},}\ }\href {\doibase 10.1103/PhysRevLett.111.064101}
  {\bibfield  {journal} {\bibinfo  {journal} {Phys. Rev. Lett.}\ }\textbf
  {\bibinfo {volume} {111}},\ \bibinfo {pages} {064101} (\bibinfo {year}
  {2013})}\BibitemShut {NoStop}%
\bibitem [{\citenamefont {Senyange}\ \emph {et~al.}(2018)\citenamefont
  {Senyange}, \citenamefont {Many~Manda},\ and\ \citenamefont
  {Skokos}}]{SMS2018}%
  \BibitemOpen
  \bibfield  {author} {\bibinfo {author} {\bibfnamefont {B.}~\bibnamefont
  {Senyange}}, \bibinfo {author} {\bibfnamefont {B.}~\bibnamefont
  {Many~Manda}}, \ and\ \bibinfo {author} {\bibfnamefont {Ch.}\ \bibnamefont
  {Skokos}},\ }\bibfield  {title} {\enquote {\bibinfo {title} {Characteristics
  of chaos evolution in one-dimensional disordered nonlinear lattices},}\
  }\href {\doibase 10.1103/PhysRevE.98.052229} {\bibfield  {journal} {\bibinfo
  {journal} {Phys. Rev. E}\ }\textbf {\bibinfo {volume} {98}},\ \bibinfo
  {pages} {052229} (\bibinfo {year} {2018})}\BibitemShut {NoStop}%
\bibitem [{\citenamefont {Many~Manda}\ \emph {et~al.}(2020)\citenamefont
  {Many~Manda}, \citenamefont {Senyange},\ and\ \citenamefont
  {Skokos}}]{MSS2020}%
  \BibitemOpen
  \bibfield  {author} {\bibinfo {author} {\bibfnamefont {B.}~\bibnamefont
  {Many~Manda}}, \bibinfo {author} {\bibfnamefont {B.}~\bibnamefont
  {Senyange}}, \ and\ \bibinfo {author} {\bibfnamefont {Ch.}\ \bibnamefont
  {Skokos}},\ }\bibfield  {title} {\enquote {\bibinfo {title} {Chaotic
  wave-packet spreading in two-dimensional disordered nonlinear lattices},}\
  }\href {\doibase 10.1103/PhysRevE.101.032206} {\bibfield  {journal} {\bibinfo
   {journal} {Phys. Rev. E}\ }\textbf {\bibinfo {volume} {101}},\ \bibinfo
  {pages} {032206} (\bibinfo {year} {2020})}\BibitemShut {NoStop}%
\bibitem [{\citenamefont {Kopidakis}\ \emph {et~al.}(2008)\citenamefont
  {Kopidakis}, \citenamefont {Komineas}, \citenamefont {Flach},\ and\
  \citenamefont {Aubry}}]{KKFA2008}%
  \BibitemOpen
  \bibfield  {author} {\bibinfo {author} {\bibfnamefont {G.}~\bibnamefont
  {Kopidakis}}, \bibinfo {author} {\bibfnamefont {S.}~\bibnamefont {Komineas}},
  \bibinfo {author} {\bibfnamefont {S.}~\bibnamefont {Flach}}, \ and\ \bibinfo
  {author} {\bibfnamefont {S.}~\bibnamefont {Aubry}},\ }\bibfield  {title}
  {\enquote {\bibinfo {title} {Absence of wave packet diffusion in disordered
  nonlinear systems},}\ }\href {\doibase 10.1103/PhysRevLett.100.084103}
  {\bibfield  {journal} {\bibinfo  {journal} {Phys. Rev. Lett.}\ }\textbf
  {\bibinfo {volume} {100}},\ \bibinfo {pages} {084103} (\bibinfo {year}
  {2008})}\BibitemShut {NoStop}%
\bibitem [{\citenamefont {Pikovsky}\ and\ \citenamefont
  {Shepelyansky}(2008)}]{PS2008}%
  \BibitemOpen
  \bibfield  {author} {\bibinfo {author} {\bibfnamefont {A.~S.}\ \bibnamefont
  {Pikovsky}}\ and\ \bibinfo {author} {\bibfnamefont {D.~L.}\ \bibnamefont
  {Shepelyansky}},\ }\bibfield  {title} {\enquote {\bibinfo {title}
  {Destruction of {A}nderson localization by a weak nonlinearity},}\ }\href
  {\doibase 10.1103/PhysRevLett.100.094101} {\bibfield  {journal} {\bibinfo
  {journal} {Phys. Rev. Lett.}\ }\textbf {\bibinfo {volume} {100}},\ \bibinfo
  {pages} {094101} (\bibinfo {year} {2008})}\BibitemShut {NoStop}%
\bibitem [{\citenamefont {Milovanov}\ and\ \citenamefont
  {Iomin}(2012)}]{MI2012}%
  \BibitemOpen
  \bibfield  {author} {\bibinfo {author} {\bibfnamefont {A.~V.}\ \bibnamefont
  {Milovanov}}\ and\ \bibinfo {author} {\bibfnamefont {A.}~\bibnamefont
  {Iomin}},\ }\bibfield  {title} {\enquote {\bibinfo {title}
  {Localization-delocalization transition on a separatrix system of nonlinear
  {S}chr\"odinger equation with disorder},}\ }\href {\doibase
  10.1209/0295-5075/100/10006} {\bibfield  {journal} {\bibinfo  {journal}
  {Europhys. Lett.}\ }\textbf {\bibinfo {volume} {100}},\ \bibinfo {pages}
  {10006} (\bibinfo {year} {2012})}\BibitemShut {NoStop}%
\bibitem [{\citenamefont {Laptyeva}\ \emph {et~al.}(2014)\citenamefont
  {Laptyeva}, \citenamefont {Ivanchenko},\ and\ \citenamefont
  {Flach}}]{LIF2014}%
  \BibitemOpen
  \bibfield  {author} {\bibinfo {author} {\bibfnamefont {T.~V.}\ \bibnamefont
  {Laptyeva}}, \bibinfo {author} {\bibfnamefont {M.~V.}\ \bibnamefont
  {Ivanchenko}}, \ and\ \bibinfo {author} {\bibfnamefont {S.}~\bibnamefont
  {Flach}},\ }\bibfield  {title} {\enquote {\bibinfo {title} {Nonlinear lattice
  waves in heterogeneous media},}\ }\href {\doibase
  10.1088/1751-8113/47/49/493001} {\bibfield  {journal} {\bibinfo  {journal}
  {J. Phys. A Math. Theor.}\ }\textbf {\bibinfo {volume} {47}},\ \bibinfo
  {pages} {493001} (\bibinfo {year} {2014})}\BibitemShut {NoStop}%
\bibitem [{\citenamefont {Vakulchyk}\ \emph {et~al.}(2019)\citenamefont
  {Vakulchyk}, \citenamefont {Fistul},\ and\ \citenamefont {Flach}}]{VFF2019}%
  \BibitemOpen
  \bibfield  {author} {\bibinfo {author} {\bibfnamefont {I.}~\bibnamefont
  {Vakulchyk}}, \bibinfo {author} {\bibfnamefont {M.~V.}\ \bibnamefont
  {Fistul}}, \ and\ \bibinfo {author} {\bibfnamefont {S.}~\bibnamefont
  {Flach}},\ }\bibfield  {title} {\enquote {\bibinfo {title} {Wave packet
  spreading with disordered nonlinear discrete-time quantum walks},}\ }\href
  {\doibase 10.1103/PhysRevLett.122.040501} {\bibfield  {journal} {\bibinfo
  {journal} {Phys. Rev. Lett.}\ }\textbf {\bibinfo {volume} {122}},\ \bibinfo
  {pages} {040501} (\bibinfo {year} {2019})}\BibitemShut {NoStop}%
\bibitem [{\citenamefont {Garc\'{\i}a-Mata}\ and\ \citenamefont
  {Shepelyansky}(2009)}]{GS2009}%
  \BibitemOpen
  \bibfield  {author} {\bibinfo {author} {\bibfnamefont {I.}~\bibnamefont
  {Garc\'{\i}a-Mata}}\ and\ \bibinfo {author} {\bibfnamefont {D.~L.}\
  \bibnamefont {Shepelyansky}},\ }\bibfield  {title} {\enquote {\bibinfo
  {title} {Delocalization induced by nonlinearity in systems with disorder},}\
  }\href {\doibase 10.1103/PhysRevE.79.026205} {\bibfield  {journal} {\bibinfo
  {journal} {Phys. Rev. E}\ }\textbf {\bibinfo {volume} {79}},\ \bibinfo
  {pages} {026205} (\bibinfo {year} {2009})}\BibitemShut {NoStop}%
\bibitem [{\citenamefont {Larcher}\ \emph {et~al.}(2012)\citenamefont
  {Larcher}, \citenamefont {Laptyeva}, \citenamefont {Bodyfelt}, \citenamefont
  {Dalfovo}, \citenamefont {Modugno},\ and\ \citenamefont
  {Flach}}]{LLBDMF2012}%
  \BibitemOpen
  \bibfield  {author} {\bibinfo {author} {\bibfnamefont {M.}~\bibnamefont
  {Larcher}}, \bibinfo {author} {\bibfnamefont {T.~V.}\ \bibnamefont
  {Laptyeva}}, \bibinfo {author} {\bibfnamefont {J.~D.}\ \bibnamefont
  {Bodyfelt}}, \bibinfo {author} {\bibfnamefont {F.}~\bibnamefont {Dalfovo}},
  \bibinfo {author} {\bibfnamefont {M.}~\bibnamefont {Modugno}}, \ and\
  \bibinfo {author} {\bibfnamefont {S.}~\bibnamefont {Flach}},\ }\bibfield
  {title} {\enquote {\bibinfo {title} {Subdiffusion of nonlinear waves in
  quasiperiodic potentials},}\ }\href {\doibase 10.1088/1367-2630/14/10/103036}
  {\bibfield  {journal} {\bibinfo  {journal} {New J. Phys.}\ }\textbf {\bibinfo
  {volume} {14}},\ \bibinfo {pages} {103036} (\bibinfo {year}
  {2012})}\BibitemShut {NoStop}%
\bibitem [{\citenamefont {for High Performance Computing (CHPC)~of
  South~Africa}()}]{CHPC2022}%
  \BibitemOpen
  \bibfield  {author} {\bibinfo {author} {\bibfnamefont {Center}\ \bibnamefont
  {for High Performance Computing (CHPC)~of South~Africa}},\ }\href@noop {}
  {}\bibinfo {howpublished} {\url{https://www.chpc.ac.za/}}\BibitemShut
  {NoStop}%
\end{thebibliography}%

\end{document}